%% file: main.tex
\documentclass[preprint,12pt]{elsarticle}

\usepackage{amssymb}

\usepackage{lineno}
\usepackage[hidelinks]{hyperref}
\hypersetup{colorlinks,bookmarksopen,linkcolor=blue,citecolor=blue,urlcolor=blue}

\journal{Elsevier}

\include{customcommands}

\begin{document}

\begin{frontmatter}

\title{Third medium finite element contact formulation for pneumatically actuated systems}

\author[ctumech,tuemom]{Ondřej Faltus\corref{cor1}}
\ead{ondrej.faltus@cvut.cz}
\cortext[cor1]{Corresponding author} 
\author[ctumech,casita]{Martin Horák}
\ead{martin.horak@cvut.cz}
\author[ctumech]{Martin Doškář}
\ead{martin.doskar@cvut.cz}
\author[tuemom]{Ondřej Rokoš}
\ead{o.rokos@tue.nl}

\affiliation[ctumech]{
            organization={Department of Mechanics, Faculty of Civil Engineering, Czech Technical University in Prague},
            addressline={Thákurova 7}, 
            postcode={166 29},
            city={Prague 6},
            country={Czech Republic}}

\affiliation[tuemom]{
            organization={Mechanics of Materials, Department of Mechanical Engineering, Eindhoven University of Technology},
            addressline={P.O. Box 513}, 
            postcode={5600MB},
            city={Eindhoven},
            country={The Netherlands}}

\affiliation[casita]{
            organization={Institute of Information Theory and Automation, Czech Academy of Sciences},
            addressline={Pod Vodárenskou věží 4}, 
            postcode={182 00},
            city={Prague 8},
            country={Czech Republic}}

\begin{abstract}
Mechanical metamaterials are artificially engineered microstructures that exhibit novel mechanical behavior on the macroscopic scale. Active metamaterials can be externally controlled. For instance, pneumatically actuated metamaterials can change their mechanical, acoustic, or other types of effective behavior in response to applied pressure with possible applications ranging from soft robotic actuators to phononic crystals. To facilitate the design of such pneumatically actuated metamaterials and structures by topology optimization, a robust way of their computational modeling, capturing both pneumatic actuation of internal voids and internal contact, is needed. Since voids in topology optimization are often modeled using a soft material model, the third medium contact formulation lends itself as a suitable stepping stone. In this manuscript we propose a single hyperelastic material model capable of (i) maintaining exactly a prescribed hydrostatic Cauchy stress within a void in the pre-contact phase while (ii) simultaneously acting as a third medium to enforce frictionless contact, which contrasts existing third medium approaches focused typically solely on contact. To achieve this goal, we split the overall third-medium energy density into contact, regularization, and pneumatic pressure contributions, all of which can be individually controlled and tuned. To prevent distortions of the compliant third medium, we include curvature penalization in our model. This improves on existing formulations in terms of compliant third medium behavior, leading ultimately to better numerical stability of the solution. Since our formulation is energetically consistent, we are able to employ more advanced finite element solvers, such as the modified Cholesky algorithm to detect instabilities. We demonstrate the behavior of the proposed formulation on several examples of traditional contact benchmarks, including a standard patch test, and validate it with experimental measurement.
\end{abstract}


\end{frontmatter}

\include{content}

\appendix
\setcounter{figure}{0}   
\include{appendices}

\bibliographystyle{elsarticle-num} 
\bibliography{biblio2.bib}

\end{document}

%% file: customcommands.tex
\usepackage[english]{babel}
\usepackage[utf8x]{inputenc}
\usepackage[T1]{fontenc}
\usepackage{bm}

\usepackage{pgfplotstable}
\usepackage{pgfplots}
\usepackage{tikz}
\usepackage{stanli}
\usepackage{amsmath}
\usepackage{amsfonts}
\usepackage{url}
\pgfplotsset{compat = 1.15}
\usepackage{graphicx}
\usepackage{color}
\usepackage{geometry}
\usepackage{setspace}
\usepackage{siunitx}
\usepackage{multirow}
\usepackage{graphicx}
\usepackage{changepage}
\usepackage{chngcntr}
\usepackage{float}
\usepackage{placeins}

\newcommand{\cof}[1]{{\rm cof} #1}

\newcommand{\dOmega}{\;\mathrm{d}\Omega}

%% file: content.tex
\section{Introduction}

The concept of a metamaterial has its origins in electromagnetic and optical applications \cite{Ren2018, Ramakrishna2005emmetamatreview}, being defined as a man-engineered material with properties pushed beyond the limits of materials that typically appear in nature. Mechanical metamaterials specifically focus on artificially designed microstructures leading to interesting mechanical characteristics on the macro scale \cite{Lee2012micronanometamatreview}, such as auxeticity (negative Poisson ratio) \cite{Ren2018, Lakes1987negativepoisson}, variable stiffness \cite{yu2018metamatreview} and programmable motion \cite{Goswami2019, Yang2015Bertoldigripper}.

Pattern-forming 2D metamaterials are a specific subclass of mechanical metamaterials originated from the behavior of a two-dimensional polymer sheet perforated in regular intervals by voids. These materials are based on a process of internal in-plane buckling of ligaments between the voids, which leads to the development of an internal pattern modifying the macroscopic behavior of the microstructure. The most prominent among these materials are honeycomb microstructures with circular voids in a hexagonal arrangement, which buckle into three different patterns depending on the biaxiality ratio of the applied macroscopic load \cite{papka1999honeycombcrushing,papka1999honeycombcrushinganalysis, Ohno2002homogenizationonhoneycombs}, or square-stacked void microstructures \cite{Mullin2007patterning}, which exhibit auxeticity in compression~\cite{Bertoldi2010auxeticityinpatterning}, possibly even in a programmable manner \cite{Florijn2014}.

The field of mechanical metamaterials is increasingly focused on design of actuated microstructures. Active control of metamaterials can be achieved by a variety of methods, such as electromagnetic or pneumatic actuation \cite{Xiao2020activereview}. Possible applications of pneumatic actuation are wide-ranging from changes in electromagnetic properties of sandwiched material designs \cite{su2020dualbandpneumaticabsorber, khodasevych2012reconfigurablefishnet} to soft robots capable of walking motion \cite{Matia2023pressureinsoftroboticactuators}. For pattern-forming metamaterials, pneumatic actuation has been studied by Chen et al. \cite{Chen2018pneumaticpatterns} to describe the relationship between the actuation and the patterning behavior. The use of a simple actuated microstructure as a gripper with programmable motion has been demonstrated by Yang et al. \cite{Yang2015Bertoldigripper}.

The natural design progression leads from intuition-based designs towards the use of generative methods such as topology optimization, which traditionally parameterizes material distribution via a density variable \cite{bendsoe2004TopologyOptimization}. This shift in the design paradigm thus necessitates a robust simulation method capable of capturing pneumatic actuation as well as internal contact in the voids upon closing, enabling multi-switching designs exploiting contact in the microstructure \cite{Coulais2018selfcontactpathways}. Even though these requirements can be met with traditional contact methods and simulation of pressure by follower loads \cite{hammer2000topoptusingforces}, such a formulation leads to a significant increase in computational complexity due to the need to identify interfaces and their normal vectors (not known a priori) from the density variable. To address this issue, topology optimization of pneumatically actuated structures performed recently by Caasenbrood et al.~\cite{Caasenbrood2020pneumatictopology} incorporated pneumatic actuation in the form of a third medium, in which an arbitrary eigenstrain is prescribed to regions with a low value of the density variable, leading to an approximate simulation of a pneumatically actuated void.

The concept of a third-medium method more traditionally appears as an alternative to classical methods for contact, such as penalty or Lagrange multiplier methods \cite{Wriggers2006contactbook}. The third medium formulation relies on the introduction of a virtual material, the so-called third-medium material, into the space between the contacting solid bodies. Properties of this material need to be chosen such that penetration is prevented without significantly influencing the solution. Among the advantages of this approach are the simplicity of meshing without the need to define contact interfaces and universal enforcement of contact regardless of which parts of the surface come into contact with each other, which is especially convenient for topology optimization. Inclusion of a third medium contact method is generally straightforward for tasks where a void space is already meshed. For instance, Dev et al.~\cite{Dev2023Electroactivefreespace} demonstrated that meshing free space is necessary for certain tasks of topology optimization of electroactive polymers. Disadvantages of the third medium method may include additional computational expenses due to meshed voids, overt reliance on material parameters of both bulk and third medium, and the need for additional regularization to prevent excessive distortions within void regions.

The third medium method for handling computational contact mechanics was first proposed by Wriggers et al. \cite{Wriggers2013thirdmedium}, even though the fictitious domain of Pagano and Alart \cite{Pagano2008fictdomain} might be considered its precursor. Wriggers et al.'s approach is largely influenced by traditional contact methods, as the third medium here is modeled as a material with variable anisotropy, the direction of which changes with the direction of the contact normal identified from the direction of the principal stretch. Despite the promising behavior demonstrated in \cite{Wriggers2013thirdmedium}, further research into the area of third medium contact remains, to the authors' knowledge, sparse. Among the rare follow-ups, we mention the work by Bog~et~al.~\cite{bog2015tmcontactwithbarrier}, and an extension to an isogeometric formulation by Kruse et al. \cite{Kruse2018isogeomthirdmedium}. Recently, a related concept has emerged that revolves around the use of an Eulerian framework to handle solid-to-solid contact~\cite{Lorez2024Euleriancontact}, which is tailored towards easy inclusion of fluid-solid interactions and growth phenomena.

The appealing applicability of the third medium contact for topology optimization was recently showcased by Bluhm et al. \cite{Bluhm2021contact}. Relying on the same hyperelastic material model for bulk material as well as voids, the authors proposed a simplified version of a third medium material with a regularization based on a locally-computed second gradients of displacement at the element level to prevent oscillations and excessive distortions of the third medium. A follow-up work with further improvement was recently presented by Frederiksen et al. \cite{Frederiksen2023topologyopt}. The design principles behind both models inspired the model of the presented in this paper.

The main contribution of this manuscript lies in integrating third medium contact with pneumatic actuation and improving on the aforementioned approaches in several ways: First, we pair an improved third medium regularization method with precise pneumatic actuation. The second-gradient regularization employed in our approach avoids unnecessarily penalizing stretch components of the deformation gradient, leading to an overall more compliant behavior of the third medium. The pneumatic actuation is represented exactly as a prescribed Cauchy stress in the material voids. Moreover, our formulation is energetically consistent. Within the present paper, we limit the scope to two-dimensional plane strain analysis, as is traditional for pattern-forming metamaterials. We comment however, on possible challenges related to extension to 3D.

The remainder of this manuscript is structured as follows. Section \ref{sec:model} provides an overview of the model with its component terms illustrated by examples throughout. Section \ref{sec:examples} then presents additional numerical examples to showcase the workings of the method, while Section \ref{sec:conclusions} offers concluding remarks together with an outline of possible future work.

\section{Proposed model}
\label{sec:model}

The desired third medium model is based on hyperelasticity and tackles both pneumatic actuation and contact. Its strain energy density function $W$ can be expected to contain three distinct terms:

\begin{equation}
    W = \psi_\mathrm{p} W_\mathrm{p} + \psi_\mathrm{c} W_\mathrm{c} + \psi_\mathrm{r} W_\mathrm{r}
\label{eq:modelenergyoverview}
\end{equation}

\noindent where the first energy term $W_\mathrm{p}$ ensures that the hydrostatic Cauchy stress across the third medium material is exactly equal to a prescribed pressure value. The neo-Hookean hyperelastic term $W_\mathrm{c}$ is then intended to enforce contact by its rapid response to material volume approaching zero, ensuring sudden stiffening of the initially compliant third medium. As pointed out in literature \cite{Bluhm2021contact}, this term is not sufficient on its own; an additional regularization term $W_\mathrm{r}$ based on second gradients of displacement is necessary to stabilize the response of the compliant third medium, especially in cases in which a free surface of the third medium is present. The magnitude of the different strain energy density terms is governed by the corresponding scalar multipliers~$\psi_\bullet$, $\bullet \in \{p$, $c$, $r\}$. In the sections below, we discuss the rationale and derivation of individual regularization terms; the pressure actuation is discussed in Section \ref{sec:model_pneumterm}, contact in Section \ref{sec:model_contactterm}, whereas the second-gradient regularization is detailed in Section \ref{sec:model_regularization}.

\subsection{Pneumatic term}
\label{sec:model_pneumterm}

Pneumatic actuation in a microstructural void means an introduction of a uniform pressure acting on the boundary of the void. The magnitude of this pressure is externally prescribed as a loading parameter. While naturally, this would be simulated by deformation-dependent loading, it can also be performed using a third medium.

This method has recently been discussed by Caasenbrood et al. \cite{Caasenbrood2020pneumatictopology} within a context of topology optimization in soft robotics. The geometric shape of a structure in topology optimization is dictated by the density variable $\rho$, which interpolates between a bulk material ($\rho = 1$) and a void ($\rho = \varepsilon$, where $\varepsilon$ is a small number, e.g. $\varepsilon = 10^{-3}$). Finite elements in regions with vanishing density variable $\rho = \varepsilon$ are selected by Caasenbrood et al. to represent pneumatically actuated voids. To introduce internal pressure, the global force vector is enhanced for each of those elements with
\begin{eqnarray}
    \bm{t}_e = \int_{\Omega_e} \mathsf{B}_{NL}^T \mathsf{D}_e \mathsf{e}_\mathrm{v} \dOmega
\label{eq:caasenbroodpneumatics}
\end{eqnarray}
\noindent where $\Omega_e$ is the element domain in 2D, $\mathsf{B}_{NL}$ is the nonlinear strain-displacement matrix that
relates displacements to the Green-Lagrange strain, and $\mathsf{D}_e$ is a constitutive elasticity tensor in the matrix format calculated using Yeoh strain energy density in \cite{Caasenbrood2020pneumatictopology}. Finally, $\mathsf{e}_\mathrm{v}$ is a prescribed constant volumetric strain that is to be interpreted as the Green-Lagrange strain. The product $\mathsf{D}_e \mathsf{e}_\mathrm{v}$ can therefore be interpreted as the volumetric part of the second Piola-Kirchhoff stress. Even though this approach may seem appropriate, it is not exactly consistent with the pressure loading, as shown below.

Herein we propose a new formulation that captures the pressure loading exactly. To this end, we introduce the following strain energy density term of the pneumatic part $W_\mathrm{p}$:
\begin{eqnarray}
    W_\mathrm{p} = pJ(\bm{F})
\label{eq:pneumaticenergy}
\end{eqnarray}
\noindent where the prescribed hydrostatic pressure $p$ works on the volumetric change $J = \det{\bm{F}}$, with $\bm{F} = \bm{I} + \nabla\bm{u}$ being the deformation gradient tensor, $\bm{I}$ the second order identity tensor, $\nabla$ the material gradient, and $\bm{u}$ the displacement vector. Integrating the strain energy density over the third medium domain $\Omega$ gives the strain energy
\begin{equation}
\Theta_\mathrm{p} (\bm{u}) = \int_\Omega \underbrace{p(\bm{x}){\rm det}(\nabla\bm{x})}_{W_p} {\rm d \Omega}    
\end{equation}
\noindent For the sake of generality and to demonstrate the full consistency of our approach with the follower force approach, we assume temporarily that the prescribed pressure $p$ can vary in the spatial coordinates $\bm{x} = \bm{X} + \bm{u}$. Taking the first variation of the strain energy and exploiting integration by parts leads to
\begin{eqnarray}
\delta \Theta_\mathrm{p} \left(\bm{u}, \delta\bm{u}\right) &=& 
  \int_\Omega J\nabla_xp \cdot\delta\bm{u}+p \;{\rm cof}\bm{F}: \nabla\delta\bm{u} {\rm d \Omega}
  \\ \nonumber
  &=& \left[\int_A\left( p  \;{\rm cof}\bm{F}\cdot\bm{N}\right) \cdot\delta\bm{u}\;{\rm d}A  +  \int_\Omega \left(J\nabla_x p -{\rm Div} \left(p {\rm cof}\bm{F}\right)\right) \cdot\delta\bm{u}\;{\rm d \Omega}\right]
\end{eqnarray}
\noindent where $\delta \bm{F} = \delta(\bm{I} + \nabla \bm{u}) = \nabla \delta \bm{u}$ and we have used the fact that the derivative of the determinant of $\bm{F}$ is its cofactor. In addition, we have introduced the unit normal in the reference configuration $\bm{N}$, the reference configuration boundary area $A$, spatial gradient $\nabla_x$, material divergence ${\rm Div}$, and the variation of the displacement field $\delta\bm{u}$. Moreover, using Nanson's formula establishing a relationship between the normal vector in the reference configuration $\bm{N}$ and the normal vector in the deformed configuration $\bm{n}$, and Piola's identity, stating that the divergence of the cofactor is zero, we arrive at
\begin{eqnarray}
 \nonumber
 \delta \Theta_\mathrm{p} &=& \Bigg[\int_A p  \;\underbrace{\left({\rm cof}\bm{F}\bm{N} \right)\cdot\delta\bm{u}\;{\rm d}A}_{\bm{n}\cdot\delta\bm{u}\;{\rm d}a}  
 \\ \nonumber
 &+&  \int_\Omega \left(\underbrace{J\nabla p\cdot\bm{F}^{-1} - \;{\rm cof}\bm{F}\cdot\nabla p}_{\bm{0}} - p \underbrace{{\rm Div}( {\rm cof}\bm{F})}_{\bm{0}}\right) \cdot\delta\bm{u}\;{\rm d \Omega}\Bigg]
 \\
 &=&\int_a p \; \bm{n} \cdot \delta \bm{u}\;{\rm d}a  
 \label{eq:followerloadconsistency}
\end{eqnarray}
\noindent which is the well-known formula for the pressure follower load boundary term \cite{Bonet2016nonlinearcontinuumfem, Kruzik2019numericalmethods}, proving the exact representation of pneumatic pressure by our improved approach. Note that in Equation (\ref{eq:followerloadconsistency}) $a$ denotes the deformed boundary area.

To further compare the presented approach with that of Caasenbrood et al., the first Piola-Kirchhoff stress tensor $\bm{P}$ can be derived as
\begin{equation}
\bm{P} = \frac{\partial W_p}{\partial \bm{F}} = p J\bm{F}^{-T}
\end{equation}
which can be transformed to Cauchy stress $\bm{\sigma}$, yielding
\begin{equation}
\bm{\sigma} = \frac{1}{J}\bm{P}\cdot\bm{F}^T = p \bm{I}
\end{equation}
\noindent resulting in a hydrostatic Cauchy stress with magnitude $p$. From this we can also compute the second Piola-Kirchhoff stress $\bm{S}$ as
\begin{eqnarray}
\bm{S} = \bm{F}^{-1}\cdot\bm{P} = p J \bm{F}^{-1}\cdot\bm{F}^{-T}
\end{eqnarray}

It may be now apparent from Equation (\ref{eq:caasenbroodpneumatics}) that through prescribing a constant Green-Lagrange strain $\bm{\varepsilon}_v = \bm{I}$, Caasenbrood et al. \cite{Caasenbrood2020pneumatictopology} essentially propose to prescribe the hydrostatic part of the second Piola-Kirchhoff stress tensor, rather than the Cauchy stress tensor, as an arbitrary constant
\begin{equation}
 p_S = \frac{1}{3}{\rm tr}(\bm{D}_e :\underbrace{\bm{\varepsilon}_v}_{\bm{I}}) = \frac{1}{3}{\rm tr}(\bm{S}) = pJ {\rm tr}(\bm{F}^{-1}\cdot\bm{F}^{-T})
\end{equation}
For a deformation-independent constitutive tensor $\bm{D}_e$ this however does not lead to constant hydrostatic pressure $p = p_S$, but rather to a deformation-dependent relation
\begin{equation}
  p = \frac{p_S}{J {\rm tr}(\bm{C}^{-1})}
\label{eq:Caasenbroodsdifference}
\end{equation}

\noindent where $\bm{C} = \bm{F}^T\cdot\bm{F}$ is the right Cauchy-Green deformation tensor. Hence, using Equation (\ref{eq:caasenbroodpneumatics}) necessitates control of $\bm{D}_e$ such that $p$ in Equation (\ref{eq:Caasenbroodsdifference}) remains constant.

Note that in the following, we expect pressurization by air pumps. Thus, we use $\Delta p$ instead of $p$ to represent a pressure difference of the air in the void as compared to the atmospheric pressure. Since we assume undeformed voids under atmospheric pressure, $\Delta p = 0$ corresponds to the initial state. Consequently suction corresponds to negative pressure difference while inflation is a result of positive pressure difference.

The contribution of the pneumatic term to the overall energy density is weighed by the scalar parameter $\psi_\mathrm{p}$. Outside of topology optimization, this material parameter is unnecessary and should always be set to~$1$. In topology optimization, where the transition from void to bulk material is smooth and only the void portion of the domain is actuated, $\psi_\mathrm{p}$ depends on the density variable $\rho$.

It is worth noting that while the presented pneumatic energy density term of Equation (\ref{eq:pneumaticenergy}) indeed results in hydrostatic Cauchy stress within the third medium, it cannot be used alone for numerical reasons, as it leads to a material with no shear stiffness. An additional energy term is needed for regularization. Its presence then distorts the perfect representation of the Cauchy stress, hence its magnitude needs to be kept comparatively low to ensure that the activated pneumatic term dominates the material energy and the contribution of other terms is negligible.

\subsection{Contact terms}
\label{sec:model_contactterm}

The contact-enforcing part of the energy density $W_\mathrm{c}$ is based on a compressible neo-Hookean material model~\cite{rivlin1948neohookean,Pence2015compressiblehyperelasticity} that is split into its volumetric $W_\mathrm{c,vol}$ and isochoric $W_\mathrm{c,iso}$ parts as

\begin{equation}
\label{eq:modelenergyNH}
    W_\mathrm{c} = \underbrace{\ln^2{J}}_{W_\mathrm{c,vol}}
    + \underbrace{\left(J^{-2/3}I_1 - 3\right)}_\mathrm{W_{c,iso}}
\end{equation}

\noindent with $I_1 = \mathrm{tr} \left(\bm{C}\right)$ the first invariant of the right Cauchy-Green deformation tensor.

\begin{figure}[h]
    \centering
    \includegraphics[width=\linewidth]{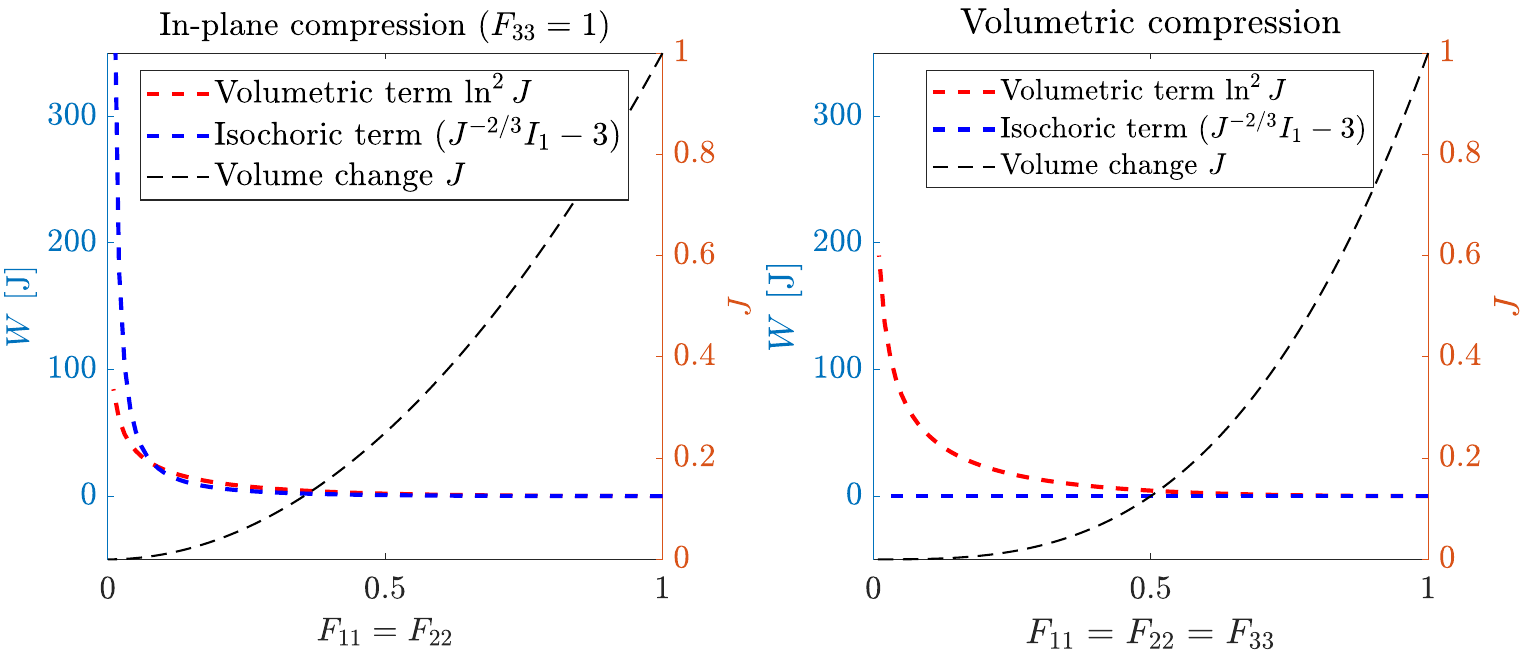}
    \caption{Behavior of the neo-Hookean strain energy density terms in response to in-plane biaxial and full triaxial compression. Under in-plane compression, due to the $J^{-2/3}$ term, the isochoric energy term tends to infinity more rapidly than the volumetric term. The isochoric term alone can therefore act as the contact term for plane strain geometries.}
    \label{fig:energies}
\end{figure}

The primary criterion for selecting a third medium contact material lies in its overall high compliance, characterized by stiffness coefficients several orders of magnitude lower than that of the bulk material constituting the contacting bodies. It is crucial, however, that initially high compliance is complemented by a swift transition to high stiffness upon total compression of the third medium, effectively preventing any form of penetration. This is satisfied, in a general 3D case, e.g., by the logarithmic nature of the volumetric term in Equation~(\ref{eq:modelenergyNH}). Under plane strain conditions, however, where the out-of-plane stretch is fixed to $1$, no purely volumetric deformation is ever possible (apart from the trivial case $\bm{F}$ = $\bm{I}$). Thus, the isochoric part by itself is active in response to in-plane compression and approaches infinity; moreover, it does so at a larger rate and more abruptly than the logarithmic term under the same conditions. This behavior is illustrated in Figure \ref{fig:energies} in comparison of plane strain and full 3D cases. As a consequence, the inclusion of the volumetric term $\ln^2 J$ of Equation~(\ref{eq:modelenergyNH}) is not strictly necessary in plane strain analysis as presented below in this paper, and hence will be left out. The volumetric part becomes necessary for 3D applications, as well as in cases where it is desirable for the third medium to correspond in its material formulation to a neo-Hookean bulk material, such as for topology optimization.

The neo-Hookean terms of the third medium model are closely related to the material model of the bulk, i.e., the contacting bodies. Their treatment differs for applications in and outside of topology optimization.

Outside of topology optimization, the material law of the bulk is arbitrary, and it is only necessary to ensure that the third medium is comparatively compliant. The material parameter $\psi_\mathrm{c}$ governing the magnitude of the neo-Hookean terms' contribution is thus set inside the third medium to a small constant positive value $\gamma$, which can be understood as the \textit{contact stiffness} of the material. The choice of $\gamma$ in relation to the material parameters of the bulk contacting bodies influences the accuracy of the solution, with larger values leading to a larger gap being left in the geometry of the converged solution. Smaller values, conversely, can lead to numerical difficulties with the solution because of badly conditioned global stiffness matrix. Natural comparison invites itself to the penalty contact method, where a similar behavior is observed: the larger the penalty parameter, the more accurate the solution and the larger the issues with stiffness conditioning. On the other hand, while the gap arising from the inaccuracy of the penalty method is negative (i.e., overlap), the gap caused by the inaccuracy of the third medium method is positive.

In topology optimization, it is advantageous to smoothly transition between bulk and void (third medium). The assignment of a region to either the void or bulk material is typically governed by the density $\rho$, i.e., a smooth field representing a gradual transition between the two materials. In a context in which the bulk material is neo-Hookean, $\psi_\mathrm{c}$ can thus be a function of the density variable or directly the density variable itself, reducing the stiffness of the base hyperelastic model in the void region and increasing it in the bulk region. Thus, a smooth transition between bulk and void is achieved. To satisfy this relationship between bulk and void, other hyperelastic material models can also supply the "neo-Hookean term" in the third medium model. The only condition is that the response to compression to zero volume needs to rapidly approach infinity, compare Figure \ref{fig:energies}.

\begin{figure}
    \centering
    \includegraphics[width=0.6\linewidth]{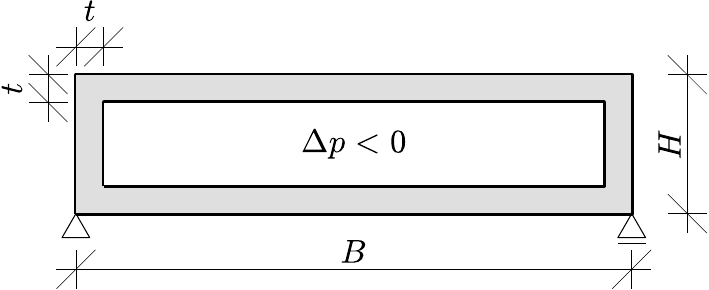}
    \caption{Pneumatic box subjected to suction within the internal void: geometry and boundary conditions.}
    \label{fig:illustration_box_geometry}
\end{figure}

The aforementioned terms represent a way to simulate pneumatic actuation with the pneumatic term, a way to stabilize the pneumatic term with the isochoric neo-Hookean term, and finally even a way to enforce contact, since the isochoric neo-Hookean term ensures that as well under plane strain conditions. With these abilities, it is now possible for the third medium model to be tested. In the following example, loosely motivated by soft robotics similarly to Caasenbrood et al. \cite{Caasenbrood2020pneumatictopology}, we are going to see the application of pneumatic suction to a rectangular box with a closed void, examining the influence of the material parameters of the third medium on the behavior of the solution. A rectangular void is positioned inside of a rectangular box, as pictured in Figure \ref{fig:illustration_box_geometry}. The dimensions of the finite sample are $B \times H = 2\;\si{\meter} \times 0.5\;\si{\meter}$ with the walls $t = \SI{0.1}{\meter}$ thick. Both bulk material and void are discretized with a regular mesh of 8-node quadratic quadrilateral elements with 9 Gauss integration points. For this and all following examples, unless specified otherwise, the bulk material is represented by a neo-Hookean solid with the bulk and shear moduli of $K = \SI{2000}{\mega\pascal}$ and $G = \SI{10}{\mega\pascal}$, respectively.

\begin{figure}
    \centering
    \begin{tabular}{c c}
         \includegraphics[width=\textwidth]{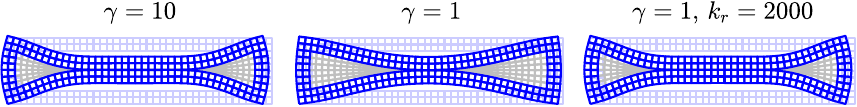} \\
         (a)
         \\
         \includegraphics[width=.6\textwidth]{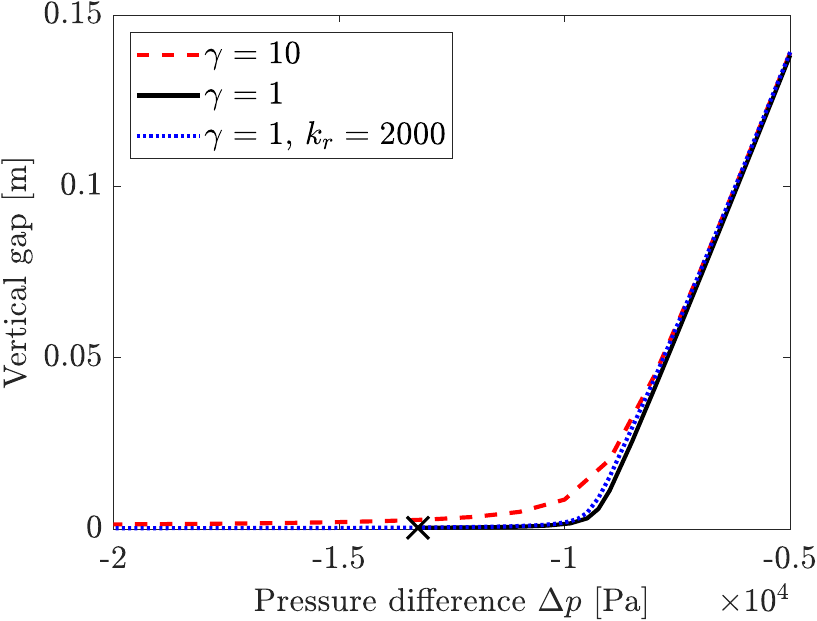}
         \\
         (b)
    \end{tabular}
    \caption{Simulation of monotone suction in a single pneumatic cell. (a)~Mesh and deformed configuration for $\gamma = 10$ (left), for $\gamma = 1$ (center), where only the last converged configuration is shown, with contact penetration and distorted third medium, and for $\gamma = 1$ with the use of additional regularizing terms (right). (b)~Comparison of pressure-gap diagrams for different values of $\gamma$ and regularization (only relevant area shown).}
    \label{fig:illustration_box}
\end{figure}

In Figure \ref{fig:illustration_box}, results of three simulations of the aforementioned example are illustrated. They differ in the parameters used for the third medium. In the first case, pictured in Figure \ref{fig:illustration_box}a (left), only the pneumatic and isochoric terms were used, with $\psi_\mathrm{c} = \gamma = 10$ and $\gamma_p = 1$. The calculation converged in the whole loading range, however this relatively large contact stiffness resulted in imprecisely identified contact onset, as can be seen from the smooth shape of the associated pressure-gap diagram in Figure \ref{fig:illustration_box}b (red dashed line). Reducing the contact stiffness $\gamma$ further to $1$ leads on the other hand to a loss of convergence of the numerical solver due to the oscillating third medium. The last converged state can be seen in Figure  \ref{fig:illustration_box}a (center) and the pressure-gap diagram cut short in Figure \ref{fig:illustration_box}b (black line with a black cross for the last converged state).

The problem of the oscillating third medium can be averted by introducing a regularization term, explained in the next section, where a new regularization stiffness parameter $k_r$ will be introduced to stabilize the third medium, see Equation~(\ref{eq:paramatersofregularization}). From the results we can already see that such a regularization leads to good accuracy while stabilizing the third medium. The satisfactory deformed shape and sharp pressure-gap diagram are visible in Figure \ref{fig:illustration_box}a (right) and from its associated blue dotted line in Figure \ref{fig:illustration_box}b, respectively. Ignoring this regularization leads either to a too stiff response or a failure of the numerical solver.

Note, in addition, that the demonstrated behavior with the third medium's computational stability depending either on sufficiently high contact stiffness at a cost of accuracy, or on additional regularization, does not only apply to pneumatic actuation; it is very similar for pure enforcement of contact conditions, too.

\begin{figure}
    \centering
    \includegraphics[width=0.6\linewidth]{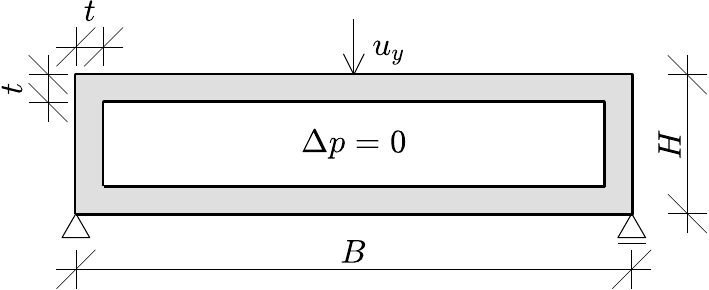}
    \caption{Geometry and boundary conditions of the second loading case, i.e., a box with an internal void subjected to displacement of its top boundary.}
    \label{fig:illustration_boxforce_geometry}
\end{figure}

\begin{figure}
    \centering
    \begin{tabular}{c c}
         \includegraphics[width=\textwidth]{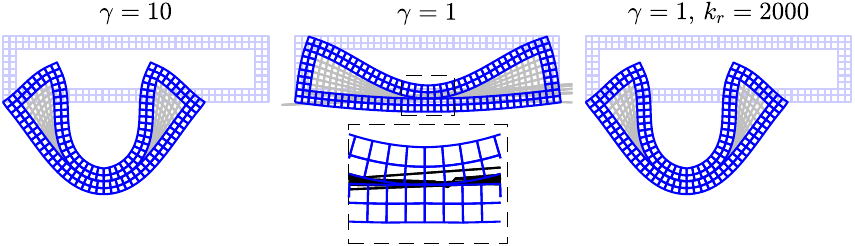} \\
         (a)
         \\
         \includegraphics[width=.6\textwidth]{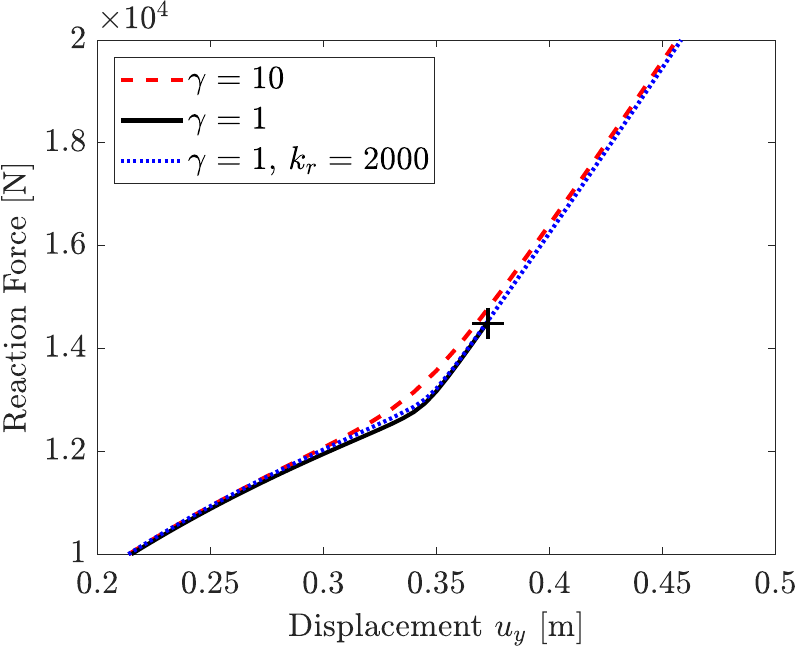}
         \\
         (b)
    \end{tabular}
    \caption{A rectangular cell loaded by a prescribed displacement, without pneumatic actuation. (a)~Mesh and deformed configuration for $\gamma = 10$ (left), for $\gamma = 1$ (center), where only the last converged configuration is shown, with contact penetration and distorted third medium (see the inset), and for $\gamma = 1$ with the use of additional regularizing terms (right). (b)~Comparison of force-displacement diagrams for different values of $\gamma$ and regularization (only vicinity of contact initiation at $u_y = 0.3\si{\meter}$ shown).}
    \label{fig:illustration_boxforce}
\end{figure}

To show this, the sample from Figure \ref{fig:illustration_box} is now loaded by a force acting in the center of the top boundary with no internal pressure applied (compare Figure \ref{fig:illustration_boxforce_geometry}), leading to a large enough deformation to enforce contact between the top and bottom walls. Results of such a simulation are pictured in Figure \ref{fig:illustration_boxforce}. For $\gamma = 10$ we can again see a smooth force-displacement diagram (red dashed line in Figure \ref{fig:illustration_boxforce}b), capturing the contact imprecisely. Reducing $\gamma$ to $1$ leads to oscillations, see Figure \ref{fig:illustration_boxforce}a center and the inset, with the force displacement diagram cut short by a loss of convergence (black line). Addition of regularization, see Section \ref{sec:model_regularization} below, solves this issue (Figure \ref{fig:illustration_boxforce}a right and blue dotted line in the force-displacement diagram).

\FloatBarrier

\subsection{Regularization term}
\label{sec:model_regularization}

The necessity of additional regularizing energy has been identified previously by Bluhm et al. \cite{Bluhm2021contact} for cases where the third medium has a free surface. However, as the previous two examples have demonstrated, regularization is needed even for enclosed void regions. Due to the generally very small stiffness of the third medium in comparison to materials of the surrounding solid bodies, it tends to deform rather excessively in the pre-contact phase. The third medium is fictional and does not describe any real material or structure, which means that this is not a problem in the physical sense; it can, however, become a problem in the computational sense. Excessive distortions limit the ability of the third medium to be volumetrically compressed while properly capturing contact. Additionally, oscillating behavior can also take place in the third medium displacements, especially for structured finite element discretizations, where multiple energetically equivalent deformations exist for the third medium.

\begin{figure}[H]
    \centering
    \includegraphics{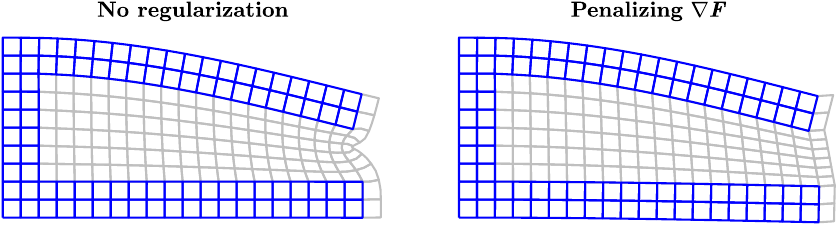}
    \caption{Effect of second gradient regularization introduced by Bluhm et al. \cite{Bluhm2021contact} on pre-contact deformation of third mediums with free surface.}
    \label{fig:noregvsbluhm}
\end{figure}

One way to ensure reasonable behavior of the third medium in the pre-contact phase is to penalize the shape distortion of its finite elements. Bluhm et~al.~\cite{Bluhm2021contact} achieved this by introducing a regularization term that penalizes second gradients of the displacement, formulating a regularizing energy term in the form of

\begin{equation}
    W_\mathrm{r} = \frac{1}{2}c\nabla \bm{F} \mathbin{\vdots} \nabla \bm{F}
\end{equation}
\noindent where $c$ is a material parameter. This leads to a much better-behaved third medium, which deforms uniformly and does not tend to bulge out of free surfaces nor oscillates, see Figure \ref{fig:noregvsbluhm} for a comparison on a simple benchmark (the detail setting of which is explained later in Section \ref{sec:bluhmc_benchmark}). A disadvantage, however, is that gradients of stretches are also penalized by this term, and thus with deformation the penalization term stiffens the third medium as a whole. While this can be addressed by recursive adjustments of material parameters, it is, in the authors' experience, rather cumbersome. Additionally, the formulation by Bluhm et al. introduces a switching term serving to restrict the regularization to compression only. Moreover, this switch, in the form of the $e^{-5J}$ multiplier, only appears in the first displacement variation of strain energy density, causing the model to become energetically inconsistent, i.e., without an energy potential. This may lead, in turn, to undesirable energy dissipation under cyclic loading and to a non-symmetric stiffness tangent. For this reason, this particular term has already been removed in a follow-up work \cite{Frederiksen2023topologyopt}.

In this contribution, we opt for a different approach based on a penalization of curvatures only, expressed as material gradients of rotations. While rotations $\bm{R}$ can be obtained from the deformation gradient $\bm{F}$ by polar decomposition $\bm{F} = \bm{R} \cdot \bm{U}$, this expression is not easily differentiable either by material coordinates $\bm{X}$ to obtain the gradients $\nabla \bm{R}$, or by deformation gradient $\bm{F}$ to calculate stresses and material stiffness necessary for any finite element implementation. A possible approach to this problem has been recently identified in the works of Smith et al. \cite{Smith2019largerotationgradients} and Poya et al. \cite{Poya2023largerotationgradients}. For the purposes of penalizing a fictitious third medium, we find it substantially easier to just penalize the gradient of an alternative rotational measure associated with the material spin, which we have denoted $\bm{Q}$ to avoid confusion with the rotation tensor $\bm{R}$. The tensor $\bm{Q}$ is defined by the incremental relation in pseudo-time $t$
\begin{eqnarray}
    \bm{Q}(t=0) = \bm{I}
    \\
    \dot{\bm{Q}} = \bm{w}_X\cdot\bm{Q}
\end{eqnarray}
\noindent where $\bm{I}$ is the second order identity tensor, $\dot{\bm{Q}}$ is the pseudo-time derivative of $\bm{Q}$, and $\bm{w}_X$ is the \textit{material spin}, defined as the skew-symmetric part of a material gradient of pseudo-velocity $\bm{v}$:
\begin{equation}
    \bm{w}_X = \frac{1}{2}\left(\nabla\bm{v} - \nabla\bm{v}^T\right) = \frac{1}{2}(\dot{\bm{F}} - \dot{\bm{F}}^T)
\end{equation}
\noindent with $\dot{\bm{F}}$ being the rate of the deformation gradient. To simplify the implementation even further, it is possible to substitute for $\bm{Q}$ its logarithm $\ln \bm{Q}$ without a significant change in effect, according to our numerical experiments. With this modification, it is now possible to construct the term $\nabla \ln \bm{Q}$ by a simple additive incremental relation. This term then serves as an easier-to-compute proxy of $\nabla \bm{R}$. Clearly, the simplifications included in the derivation of this term represent a compromise between computational convenience and physical accuracy in capturing curvature, but we believe it is acceptable due to the fictitious nature of the third medium. A detailed discussion of this reasoning and the description of the regularization term's implementation can be found in \ref{app:rotations}. The penalization thus follows from a quadratic strain energy density term

\begin{equation}
    W_r = \frac{1}{2}c \left(\nabla \ln \bm{Q} \mathbin{\vdots} \nabla \ln \bm{Q}\right)
\end{equation}

\begin{figure}[h]
    \centering
    \includegraphics{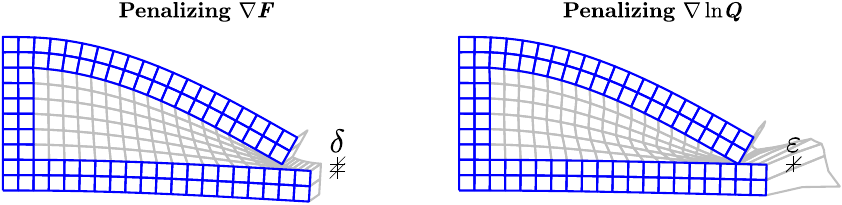}
    \caption{Comparison of the effect of regularization terms $\nabla \bm{F}$ and $\nabla \ln \bm{Q}$. The regularization based on rotation ($\nabla \ln \bm{Q}$) leads, for the same material parameters, to a lower stiffness of the third medium and hence smaller contact gap ($\varepsilon << \delta$).}
    \label{fig:bluhmvsrot}
\end{figure}

The regularization term based on the $\bm{Q}$-rotation indeed leads to element shape regularization without affecting the  volumetric stiffness of the third medium. Consider Figure \ref{fig:bluhmvsrot}, where two simulations are compared with the same material parameters on the C-shape specimen, showing the additional stiffness exerted by the original $\nabla \bm{F}$ regularization compared to the $\nabla \ln \bm{Q}$ regularization, visible in the larger contact gap.

\begin{figure}[H]
    \centering
    \includegraphics{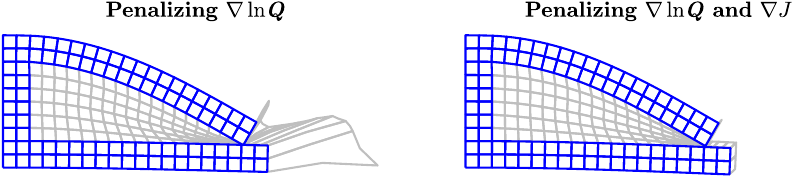}
    \caption{Comparison of the effect of regularization terms. The regularization based on rotation ($\nabla \ln \bm{Q}$, left), can lead to numerical instability due to flipped elements and hence negative finite element Jacobians. This is avoided by adding an enforcement of uniform volume change across element (penalizing $\nabla J$, right).}
    \label{fig:rotvsjesr}
\end{figure}

There is one drawback, however: due to the larger freedom in element deformation caused by the lack of stretch and shear penalization, elements may approach a deformed state with locally zero volume in certain geometries of the third medium. The resistance of the contact part of the third medium model to this condition then leads either to numerical instabilities (consider Figure~\ref{fig:rotvsjesr}) or, for stiffer material parameters, to the creation of an artificially stiff band across the geometry of the third medium and consequently premature enforcement of contact constraints. Fortunately, these issues are easily remedied by an additional penalty on the gradient of $J$, forcing uniformity of volume change across each element. Thus, the final regularization term we propose consists of a sum of two quadratic forms:

\begin{equation}
    W_\mathrm{r} = \frac{1}{2}c\left(\nabla \ln \bm{Q} \mathbin{\vdots} \nabla \ln \bm{Q} + \nabla J\cdot \nabla J\right)
\end{equation}

The proposed regularization term meets expectations and allows for regularization without influencing the contact stiffness and with stable numerical behavior. Uniform rotation, i.e., zero curvature and uniform volume change, are enforced. Contrary to Bluhm et al.'s approach of penalizing second displacement gradients as a whole, however, stretch and certain modes of symmetric extension or contraction of the elements are left unpenalized, which improves the compliant behavior of the third medium. Detailed comparison of the different approaches to regularization with numerical examples can be found in Section \ref{sec:bluhmc_benchmark}. Apart from being much less dependent on the contact stiffness, the present approach also tends to perform better with more compliant third mediums, which greatly increases the accuracy of contact constraint enforcement.

Because of this regularization term, any finite element use of the third medium model necessitates discretization into at least quadratic elements. It is also worth noting that such an implementation is local only in the rotational and Jacobian gradients, with no continuity required of the second gradients across element boundaries. For practical reasons, higher continuity is not enforced, since this would require using splines as interpolation functions \cite{Fischer2011IGA} or introduction of an additional gradient field coupled weakly with displacements \cite{Horak2020gradientpolyconvexity}. While this is a physical violation in a model of a realistic material, it is of no consequence here due to the fictitious nature of the third medium.

The material parameter $\psi_\mathrm{r}$ should reflect the contact stiffness already included in the neo-Hookean terms' multiplier. In all examples here, it is set to

\begin{equation}
    \psi_\mathrm{r} = \gamma {k}_r
\label{eq:paramatersofregularization}
\end{equation}

\noindent where $\gamma$ is the previously discussed contact stiffness appearing in $\psi_c$ and $k_r$ is a regularization stiffness coefficient set usually to a value between $10^2$ and $10^6$. To maintain unit consistency of material parameters, it is possible to include some characteristic stiffness and length parameters of the bulk material within $c$. Within topology optimization, a switch based on the density variable is necessary to disable the regularization term for bulk regions.

\section{Numerical experiments}
\label{sec:examples}

We demonstrate efficacy of the proposed third medium model on three benchmark and test problems: a patch test (Section \ref{sec:patchtest}), a self-contact example inspired by Bluhm et al. \cite{Bluhm2021contact} (Section \ref{sec:bluhmc_benchmark}), and a pneumatic actuation of a pattern-forming metamaterial sample (Section \ref{sec:metamaterial}). In the spirit of topology optimization and pattern-forming metamaterials, we use a neo-Hookean material model for the solid material, which could represent a silicone rubber sample produce by casting in a 3D-printable mold. The bulk and shear moduli $K = \SI{2000}{\mega\pascal}$ and $G = \SI{10}{\mega\pascal}$ are used. The model is defined with the strain energy density in the form

\begin{equation}
    W_\mathrm{NH,bulk} = \frac{K}{2}\ln^2{J}
    + \frac{G}{2}\left(J^{-2/3}I_1 - 3\right)
\label{eq:bulkenergy}
\end{equation}

The discretization of each problem consist of quadratic 8-node quadrangular finite elements with 9 Gauss integration points. We found that a Lobatto integration rule (i.e., integration points at element nodes) improves the precision of contact enforcement. In the authors' experience, this improvement is particularly significant for coarser meshes. With standard integration rules, interpenetration of contact surfaces might occur, its magnitude being approximately the distance between the element boundary and the integration point of the third medium element. For meshing, the open-source GMSH software has been used \cite{geuzaine2009gmsh}. All presented examples have been calculted under plane strain conditions.

For the patch test benchmark, the standard Newton solver has been used to solve the system. For the subsequent examples, a modified Newton algorithm \cite{Gill1974newtonforoptimization,Fletcher1977modifiedNewton} has been employed, using Cholesky decomposition to modify the Hessian to repel the solution from local maxima and saddle points towards a local minimum. Additionally, a smart step length reduction scheme was in effect to allow the simulation to proceed in cases where large deformations, instabilities, and oscillations within the third medium deformed geometry may occur. When the solver had not converged within a specified maximum of Newton iterations, step length was halved, this process repeating for a maximum of 14 times before concluding the solver had indeed failed. Successful convergence of a Newton step lead to an increase of step length by the factor of $1.5$, conversely.

\subsection{Patch test}
\label{sec:patchtest}

\begin{figure}
    \centering
    \includegraphics[width=0.4\linewidth]{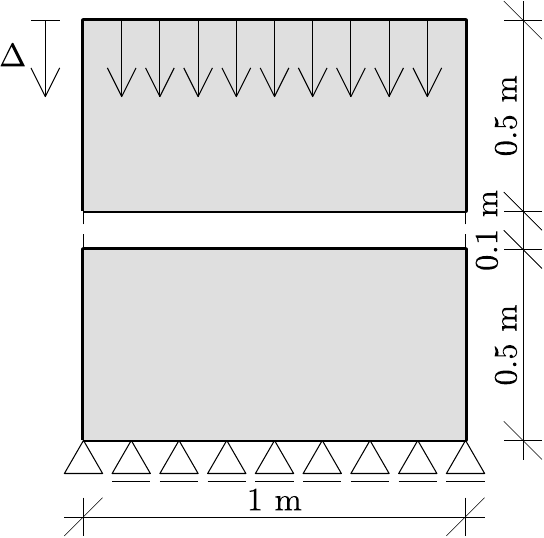}
    \caption{Geometry and boundary conditions of the standard patch test \cite{Crisfield2000patchtest} as adapted for third medium contact.}
    \label{fig:patchtest_geometry}
\end{figure}

Performance of a computational contact method is traditionally verified using a standard contact patch test. Here, we adopted a patch test proposed originally by Crisfield \cite{Crisfield2000patchtest}, cf. Figure \ref{fig:patchtest_geometry}, in which a block of an elastic material of dimensions $1 \si{\meter}\times 1 \si{\meter}$, cut in two by a horizontal contact interface, is loaded by a uniform downward displacement on the top boundary of the upper layer, thus pressing the top part against the bottom one. This should lead to a uniform stress distribution in the material, identical on both sides of the contact interface. For small strain elasticity in a two-dimensional domain under plane strain conditions, an analytical solution of $\sigma_{22} = E\Delta/(1-\nu^2)$ can be derived, with $\sigma_{22}$ being the vertical normal component of the stress tensor, $E$ the elasticity modulus, $\Delta$ the imposed displacement (corresponding to vertical strain due to the unit dimensions of the block), and $\nu$ the Poisson's ratio. If the interface nodes of the finite element mesh are aligned across the contact interface, the test is typically fulfilled for all contact methods. If they are misaligned, however, some commonly used methods may fail unless sufficiently modified, resulting in a horizontally non-uniform pressure distribution, see, e.g., \cite{Zavarise2009improvedNTS}.

For the purposes of third medium contact, a slight modification is necessary: an initial gap has to be introduced in the contact interface to accommodate a third medium mesh inbetween the two elastic blocks, see Figure \ref{fig:patchtest_geometry}. We set the gap to be $\SI{0.1}{\meter}$ high, with the blocks placed above and below the gap without being reduced in size. Total prescribed displacement on the top boundary is $\Delta = -\SI{0.2}{\meter}$; that is, $-\SI{0.1}{\meter}$ to close the gap, and $-\SI{0.1}{\meter}$ to induce a vertical strain of $\varepsilon_{22} = 0.1$ in the elastic block. Parameters of the third medium model used for this benchmark are $\gamma = 1$ and $k_r = 2\times 10^3$, recall Equation (\ref{eq:paramatersofregularization}). Since a large strain hyperelastic material is used for the bulk, instead of relying on a small-strain analytic solution, the comparison will be made with a numerical result from an undivided elastic block without any contact interface influence (i.e., a \textit{contact-free solution}).

\begin{figure}
    \centering
    \hspace{5mm}
    \begin{tabular}{c c}
         \includegraphics[width=.45\linewidth]{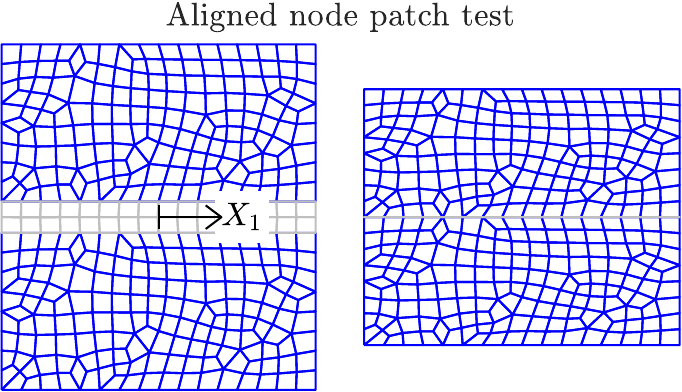}
         &
         \includegraphics[width=.45\linewidth]{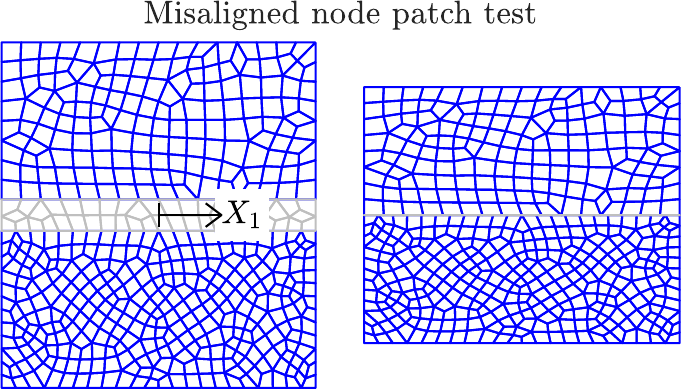}
         \\
         (a) & (b)
    \end{tabular}
    \begin{adjustwidth}{-0.1\textwidth}{-0.1\textwidth}
    \centering
    \begin{tabular}{c}
         \includegraphics[width=1.2\textwidth]{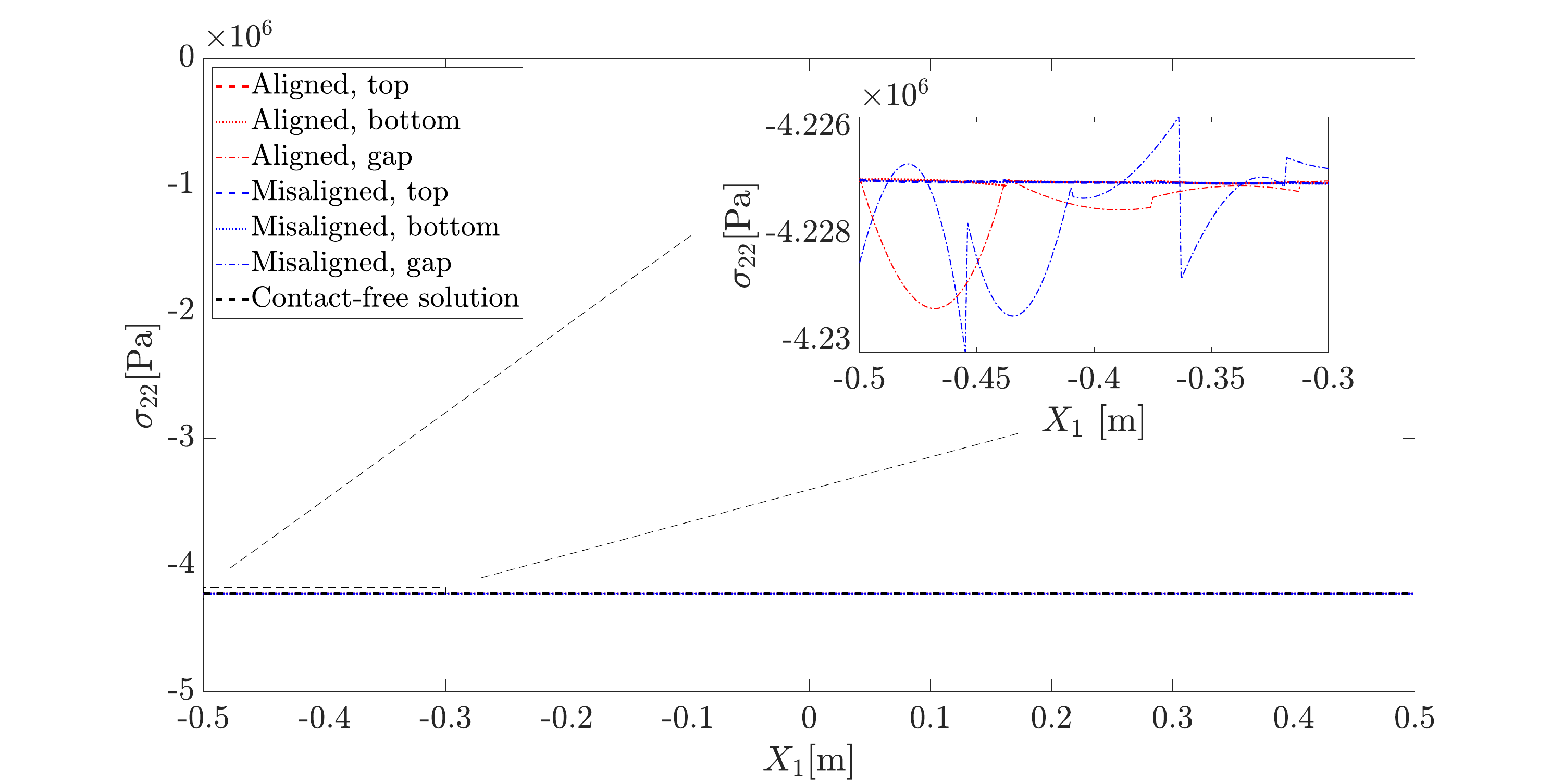}
         \\
         (c)
    \end{tabular}
    \end{adjustwidth}
    \caption{Patch tests of the presented third medium contact formulation conducted on an elastic block with a gap with uniform displacement applied to the top boundary. Illustration of the undeformed and the deformed configurations (bulk material in blue, third medium in gray) for two test variants: (a) aligned interface nodes, (b) misaligned interface nodes. (c) Uniform stress propagation through the third medium, resulting in a uniform stress distribution in the elastic block, with identical magnitudes on the top and bottom; comparison with contact-free benchmark solution. Note that a slight stress increase and oscillations around the gap near the interface are present, negligible compared to its average.} 
    \label{fig:patchtest}
\end{figure}

Figure \ref{fig:patchtest} shows results of two patch test variants, one with aligned interface nodes and the other with misaligned interface nodes. Their meshes and corresponding deformed states can be seen in Figures \ref{fig:patchtest}a-b. Figure \ref{fig:patchtest}c, combined for both cases, reveals a complete overlap among the uniform stress distributions of the contact-free solution and of the benchmark results. Those were sampled along the top and bottom boundaries of the block, as well as along the gap. The third medium method thus fulfills the patch test for both the aligned and misaligned node cases. Admittedly, a detailed look at the resulting stress distribution reveals slight imperfections of the solutions, see the inset of Figure \ref{fig:patchtest}c. Oscillations in the stress distributions in the vicinity of the gap can be observed, while all the stress curves tend to rise slightly close to the vertical edge of the domain. This effect is caused by the small free surface of the third medium on both sides of the gap, where the third medium elements have the freedom to bulge out. It can be reduced by a more aggressive second gradient regularization, i.e. by setting a larger $k_r$, or completely avoided by a boundary condition preventing horizontal movement of the free surface. Both this free surface effect and the oscillations, however, remain in magnitudes negligible compared to the solution.

\subsection{Self-contacting C-shape example}
\label{sec:bluhmc_benchmark}

\begin{figure}
    \centering
    \includegraphics[width=.5\textwidth]{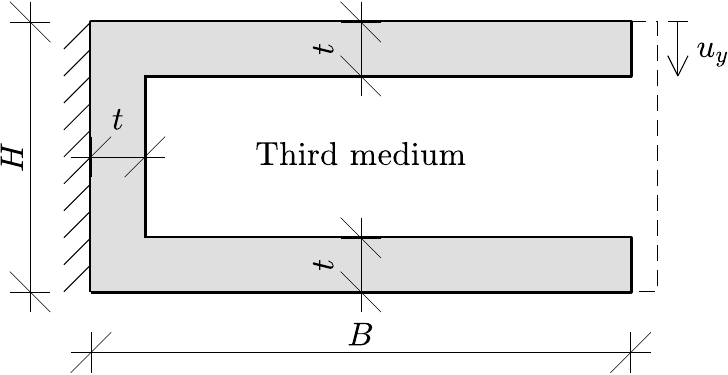}
    \caption{Geometry and boundary conditions of the C-shape benchmark example originally proposed by Bluhm et al. \cite{Bluhm2021contact}.}
    \label{fig:cshapegeom}
\end{figure}

Bluhm et al. \cite{Bluhm2021contact} have recently developed a particular benchmark for third medium contact, which has proven useful to demonstrate the characteristic behavior of third medium contact models. It consists of a rectangular structure in the shape of the letter C elongated along the horizontal axis, as pictured in Figure \ref{fig:cshapegeom}. The dimensions in the examples presented here forth are $B=\SI{1}{\meter}$, $H=\SI{0.5}{\meter}$ and $t=\SI{0.2}{\meter}$. The entire vertical left edge of the C-shape domain is fixed. The rightmost corner of the upper horizontal cantilever is loaded with a prescribed displacement $u_y$, forcing it to bend up to the point of initial contact with the lower cantilever and beyond. To capture this contact, the inner area of the C-shape specimen has to be filled with the third medium material. The third medium, therefore, has a relatively long free surface along the right-hand side vertical boundary of the domain. The particular usefulness of this simple benchmark is twofold: first, the rectangular geometry of the domain allows for perfectly regular meshes, on which all distortions of the third medium and its boundary are clearly visible; second, the contact point itself lies on the free surface of the third medium.

\begin{figure}
    \centering
    \includegraphics[width=\textwidth]{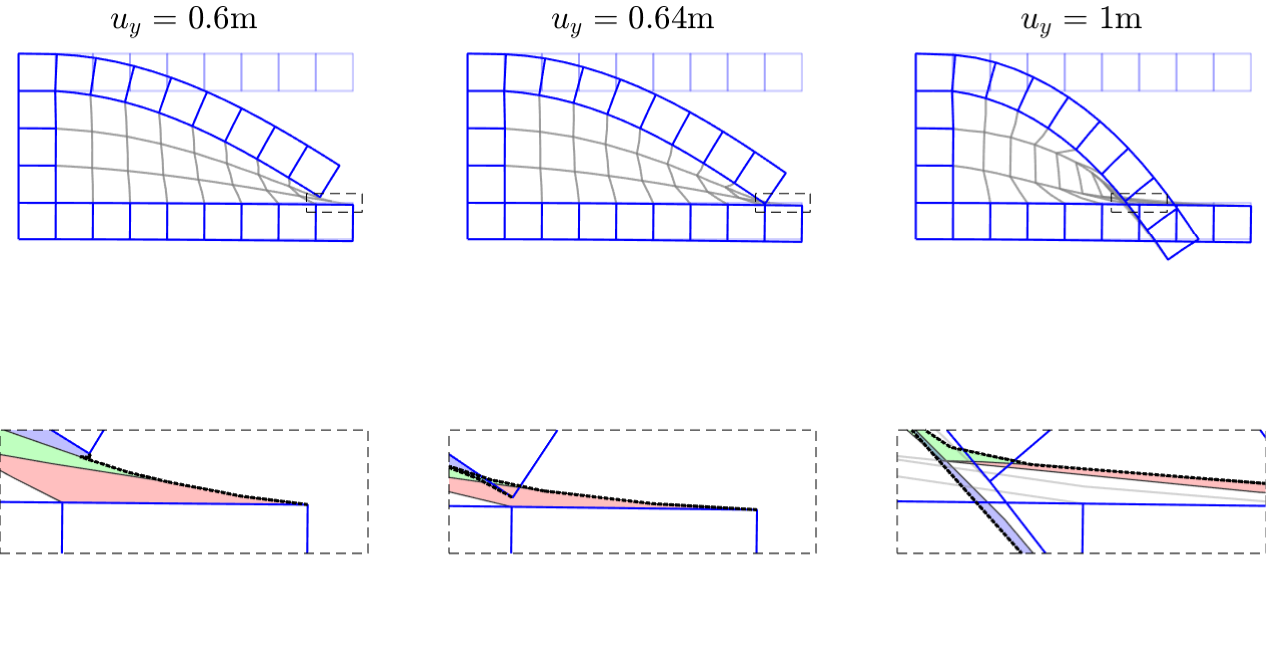}
    \caption{Results of a tentative C-shape benchmark without an extra column of third medium elements. With loading, the free surface (the black dashed line in the insets) bulges inward (left column), leading to interpenetration of non-neighboring finite elements (center column) and finally failure to enforce contact constraints (right column). Note the overlapping colored elements of the third medium in the detailed insets.}
    \label{fig:extracolumn}
\end{figure}

Among the peculiarities of this benchmark problem is the necessity to include at least one extra column of third medium elements along the free surface. This has already been demonstrated by Bluhm et al. \cite{Bluhm2021contact}, and remains true for the present material model as well. The reason behind this requirement is purely numerical, caused by the behavior of the deformed third medium free surface. This is illustrated in Figure \ref{fig:extracolumn}, where a very coarse third medium mesh, consisting only of three elements per gap height without this extra column is shown pressed up to the point of contact and beyond. If the vertical free surface bulges ever so slightly inwards, penetration occurs, with the contact point passing through the boundary of a non-neighboring third medium element. Thus, the third medium is prevented from arranging itself in the contact gap and activating the resistance to compression of the hyperelastic law.

\begin{figure}
    \centering
    \includegraphics[width=\linewidth]{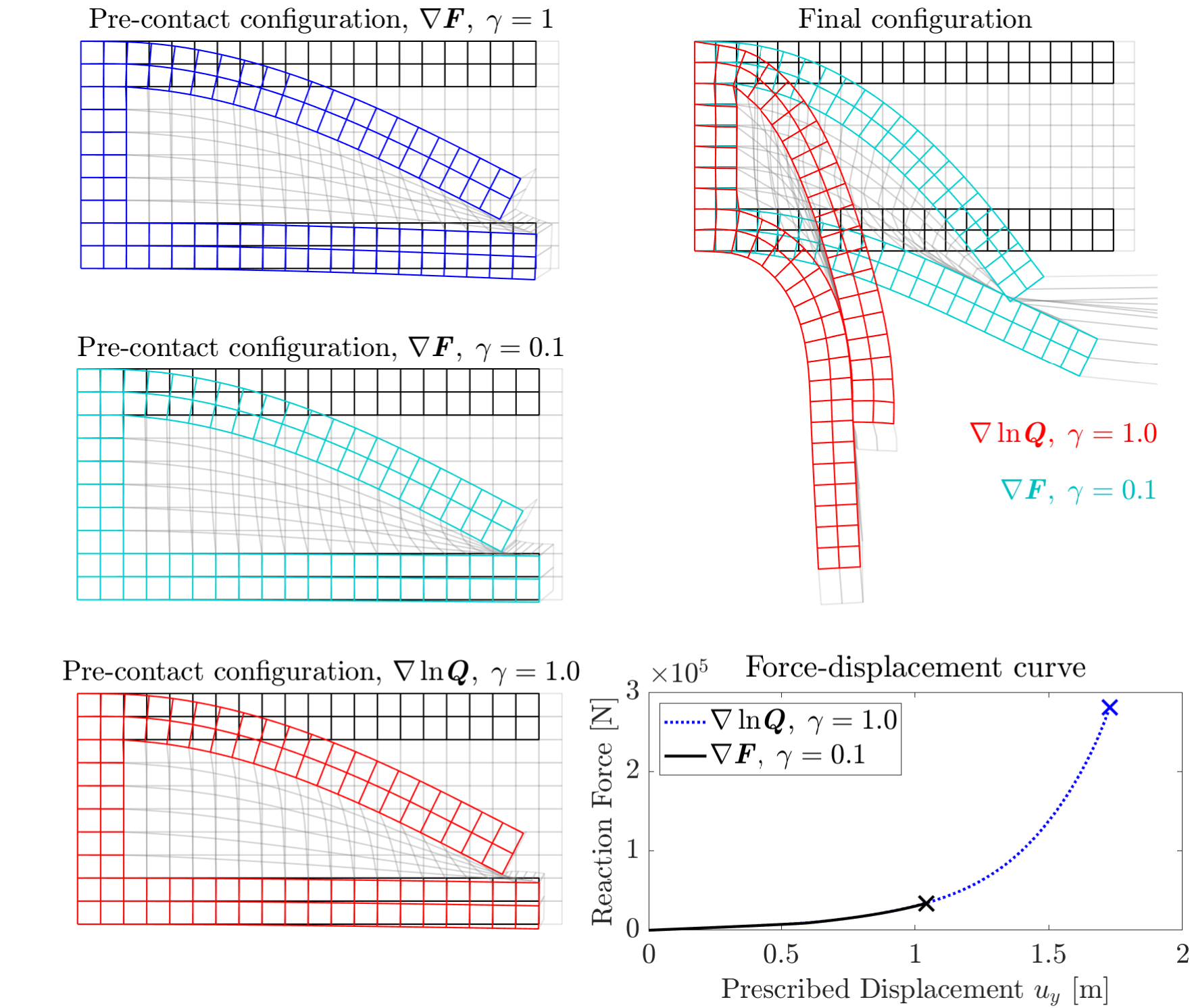}
    \caption{Self contact example and a comparison of the newly proposed  regularization $\nabla \ln \bm{Q}$ to a $\nabla \bm{F}$ formulation of Bluhm et al. \cite{Bluhm2021contact}. In the pre-contact phase, $\nabla \bm{F}$ with $\gamma = 1$ exhibits premature contact (in blue), unlike $\nabla \ln\bm{Q}$ (in red) or $\nabla \bm{F}$ with $\gamma = 0.1$ (in cyan). The final configuration plot and its associated force-displacement curves with highlighted points of failure demonstrate the higher robustness of the $\nabla \ln\bm{Q}$ regularization. The $\nabla \ln \bm{Q}$ formulation only becomes numerically unstable when the contact surface starts being loaded in a tangential direction. }
    \label{fig:cshapecomparison}
\end{figure}

A correctly behaving C-shape benchmark simulation, i.e., with the additional layer of elements, is depicted in Figure \ref{fig:cshapecomparison}, where the newly proposed approach to regularization of the third medium (denoted herein as $\nabla \ln \bm{Q}$, see Section \ref{sec:model_regularization}) is compared to the original approach of Bluhm et al. \cite{Bluhm2021contact} (denoted here as $\nabla\bm{F}$).

The parameters of the third medium for the $\nabla\ln\bm{Q}$ regularization are $\gamma = 1, k_r = 2\times 10^3$, whereas for the $\nabla\bm{F}$ regularization the regularization stiffness $k_r = 2\times10^3$ and $\gamma \in \{0.1, 1\}$ is used. Since the contact stiffness $\gamma = 1$ leads to premature contact, we include also a lower setting of $\gamma = 0.1$.

The newly proposed $\nabla \ln \bm{Q}$ regularization proves to be more robust, allowing for much larger deformation (see Figure~\ref{fig:cshapecomparison}c) before the simulation stops due to numerical oscillations and instabilities caused by large sliding along the contact surface (as the surface has rotated to be loaded almost entirely in a tangential direction). This is mainly due to the ability of the new formulation to utilize a larger value of contact stiffness without experiencing premature contact, thus achieving better performance, with the computational cost of both formulations being comparable.

\subsection{Buckling of an internally pressurized metamaterial}
\label{sec:metamaterial}

The last example showcases the third medium model in simulations of pneumatically actuated pattern-forming metamaterials. Specifically, we consider a silicone rubber sample with four circular voids arranged in a square lattice. This arrangement is known upon compression to change its geometry from initially circular holes to elliptical holes with alternating horizontal and vertical major axes as a result of an internal instability, resulting in auxetic behavior \cite{Mullin2007patterning, Bertoldi2010auxeticityinpatterning}. Alternatively to mechanical compression, this behavior can also be introduced by pneumatic pressure in the voids \cite{Chen2018pneumaticpatterns}.

\begin{figure}[h]
    \centering
    \includegraphics[width=0.4\linewidth]{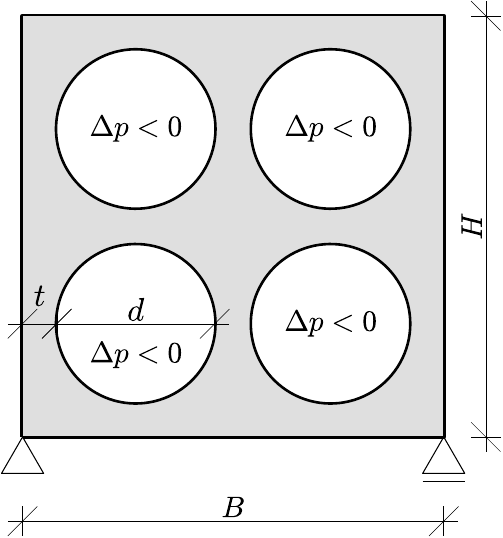}
    \caption{Geometry and boundary conditions of the symmetrical metamaterial rubber sample.}
    \label{fig:metamat_geometry}
\end{figure}

Here, the numerical model is compared against experimental data. To this end, a square specimen has been manufactured and tested with the dimensions of $B \times H = \SI{40}{\milli\meter} \times \SI{40}{\milli\meter}$ and four voids of $d = \SI{15}{\milli\meter}$ diameter located $t = \SI{3.25}{\milli\meter}$ from the edge, see Figure \ref{fig:metamat_geometry}.

\begin{figure}
    \centering
    \begin{tabular}{c}
         \includegraphics[width=.8\textwidth]{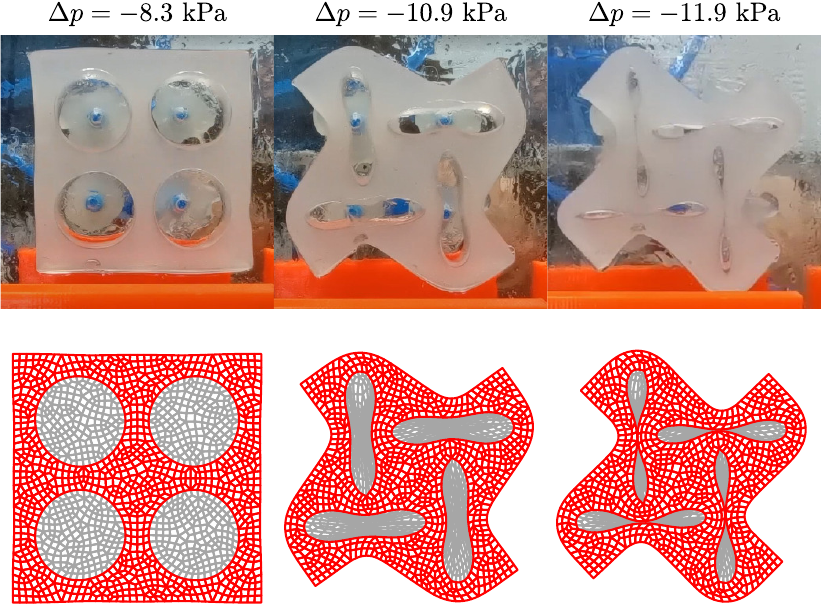}
         \\
         (a)
         \\
         \includegraphics[width=.6\textwidth]{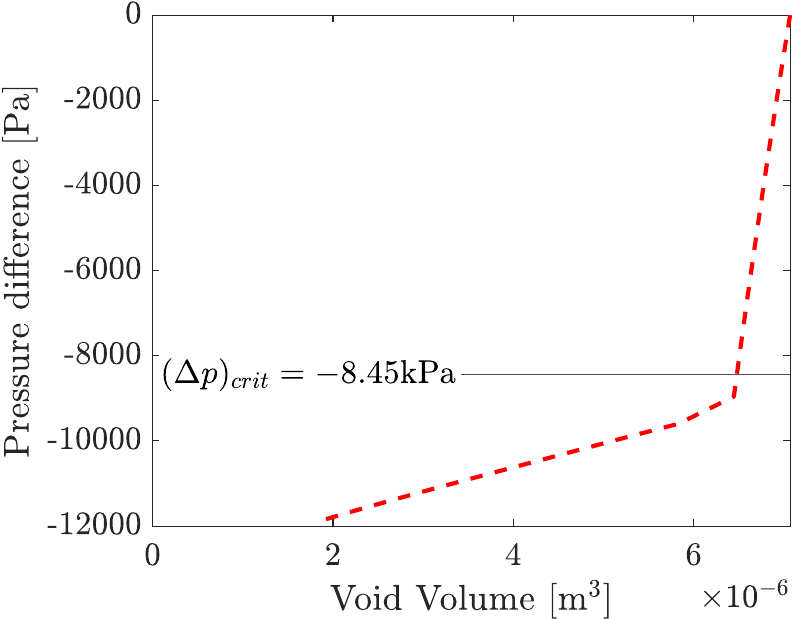}
         \\
         (b)
    \end{tabular}
    \caption{Simulation and experimental data of a pneumatically actuated sample of a pattern-forming metamaterial. (a) Comparison of the deformed states from simulation and experiment for various levels of internal pressure loading. (b) A corresponding pressure-volume diagram, with an apparent point of internal instability. The horizontal line highlights the experimentally determined buckling pressure of  $(\Delta p)_\mathrm{crit} = \SI{-8.45}{\kilo\pascal}$.}
    \label{fig:metamat}
\end{figure}

In the experiment, the sample was placed between a pair of lubricated plexiglass sheets, see Figure \ref{fig:metamat}a. Through four drilled holes therein, air pumps were attached to introduce a constant suction pressure into all four voids. The evolution of pressure difference with respect to the ambient atmospheric pressure was measured during the experiment to determine a critical value of the pressure difference leading to internal buckling. In a repeated pressure loading and unloading procedure, the sample was observed to buckle upon loading and then return to an almost undeformed configuration upon unloading. Numerical sensitivity analysis with regards to geometric imperfections shows that this effect of material memory should have a negligible influence on the results. The buckling point was consistently achieved for the critical pressure of $(\Delta p)_\mathrm{crit} = \SI{-8.45}{\kilo\pascal} \pm \SI{0.3}{\kilo\pascal}$.

To reflect the conditions of the experiment, a square-shaped finite structure with free edges except for preventing rigid body motion is considered in the numerical simulation. The considered plane strain model corresponds well to reality, since the out-of-plane deformation is prevented by the fixed pair of plexiglass sheets. For the silicone rubber, yet again the neo-Hookean model is used with its strain energy density as defined in Equation (\ref{eq:bulkenergy}). The material parameters $K$ and $G$ have been determined from an assumption of near incompressibility and uniaxial compression tests on samples of the silicone rubber, resulting in the values of $K = \SI{91.111}{\mega\pascal}$ and $G = \SI{0.182}{\mega\pascal}$. Further details of the material parameter identification are included in \ref{app:matparams}.

The third medium parameters in this case are chosen as $\gamma = 1\times10^{-3}$ and $k_r = 2\times10^3$. The $\nabla \ln \bm{Q}$ model with pneumatic actuation is used. The contact stiffness is somewhat lower than the previous examples to maintain comparative compliance of the third medium in relation to the parameters of the solid material.

In both the simulation and experiment, the negative pressure difference leads to an internal instability causing a pattern to emerge in the shape of the voids as a result of bifurcation. Photos of the experiment are compared to deformed configurations resulting from the simulation in Figure \ref{fig:metamat}a for three values of pressure difference: one pre-bifurcation, one post-bifurcation, and one at the end of the experiment where voids are in contact. The critical pressure value of the instability found experimentally is highlighted in a pressure-volume diagram in Figure \ref{fig:metamat}b. In the simulation, the critical pressure value obtained was slightly lower, namely $(\Delta p)_\mathrm{crit} = \SI{-8.96}{\kilo\pascal}$. This slight error is explained by measurement imprecision, viscous effects in the silicone rubber material not captured by the hyperelastic model, general geometrical imperfection of the experiment, and friction between the plexiglass sheets and the specimen. Nevertheless, it can be said that the numerical results agree satisfactorily with the experimental observations, with the model accurately capturing the experimentally determined behavior.

\section{Conclusions}
\label{sec:conclusions}

In this manuscript, we have introduced a novel material model for a computational fictitious third medium enabling concurrent modeling of pneumatic actuation and contact. Combined pneumatic actuation and contact can be a strong design tool in research, topology optimization, and design within the soft robotics field and mechanics of secondary stiffening or densification of pattern-forming metamaterials. The proposed model is formulated as hyperelastic, albeit with inclusion of second gradient terms. Its strain energy density function comprises three distinct parts representing the different functionality of the third medium (contact, regularization of pre-contact behavior, pneumatic pressurization). Individual parts can be switched on and off independently depending on the use case. Noteworthy is the utility of this formulation for topology optimization of both metamaterials and macrostructures. In that case, the presence of a third medium mesh is implicit, which eliminates the main disadvantage of all third medium methods: the need for extra computational effort due to additional finite elements within void regions.

The main contributions of this manuscript can be summarized as follows. Several limitations of existing third medium formulations for contact and pneumatic actuation have been eliminated. First, our pneumatic actuation model describes pressure loads precisely, maintaining consistency with a follower-load approach. Next, a second gradient regularization term, which governs the behavior of the compliant third medium, significantly improves computational stability. Our modification is based on an energy potential, meaning that the deformation of the third medium is thus fully reversible without an energy dissipation and the entire model has a symmetric tangent stiffness. This further allows for the employment of advanced mathematical programming methods and finite element solvers, such as buckling analysis. Finally, only material gradients of selected deformation mode fields are penalized, greatly improving the compliant behavior of the third medium.

The proposed methodology has been demonstrated on a number of numerical examples, demonstrating the usefulness of the model for both free surface and closed void configurations including simulations of pneumatically actuated structures. In all cases, our formulation exhibits in improved behavior compared to existing third medium contact formulations. It is furthermore shown that presented contact method fulfills the standardized benchmark patch test.

Although the model formulation has been presented for plane strain geometries, its extension to 3D is straightforward, albeit technical. The incremental additive method for the calculation of rotational gradient terms relies on commutation between rotation tensors, generally not present in 3D geometries; the gradient terms can, however, be reformulated in a more general exponential form to avoid this limitation. Desired future extensions concern implementation within a topology optimization framework and inclusion of frictional contact and, which is an ongoing effort \cite{Wriggers2013thirdmedium}.

\paragraph{Acknowledgement} This work has been supported by the Czech Science Foundation [grant no.~GA19-26143X]. The authors would like to thank Jan Havelka and Pavel Rychnovský at the Czech Technical University in Prague for their help with physical experiments.

%% file: appendices.tex
\section{Implementation of second gradient regularization}
\label{app:rotations}

Third medium is prone to excessive deformation of its highly compliant finite elements, which can lead to numerical instabilities. To prevent this issue, it is necessary to penalize sharp changes in rotations, i.e., to penalize curvatures. To this effect, a regularization term is introduced, penalizing the gradient of large rotations. The rotation tensor $\bm{R}$ can be obtained by polar decomposition of the deformation gradient $\bm{F}$ into stretch and rotations as
\begin{equation}
    \bm{F} = \bm{R} \cdot  \bm{U}
\end{equation}
However, calculation of $\nabla \bm{R}$ is computationally challenging, requiring spectral decomposition of $\bm{F}$, which is process that is difficult to differentiate. Since the desired penalization is not physical but purely numerical, we can opt for penalizing related quantities instead. Thus we propose to penalize the gradient of the logarithm of an alternative rotational measure $\bm{Q}$, i.e. $\nabla \ln \bm{Q}$, which can be in the given plane strain finite element context, calculated more easily with the use of incremental linearization.

The $\bm{Q}$-rotations can be incrementally calculated in pseudo-time derivative from the material spin tensor $\bm{w}_X$ as
\begin{equation}
\label{eq:Qtimederivative}
    \dot{\bm{Q}} = \bm{w}_X \cdot \bm{Q}
\end{equation}
The material spin tensor $\bm{w}_X$ is defined as the skew-symmetric part of the \emph{material} gradient of the spatial velocity $\nabla\bm{v}$:
\begin{equation}
\label{eq:spin}
    \bm{w}_X = \frac{1}{2} \left(\nabla\bm{v} - \left(\nabla\bm{v}\right)^T\right) 
\end{equation}
\noindent The spatial velocity $\bm{v}$ is defined as the change of displacement $\bm{u}$ within a computational timestep~$\Delta t$:
\begin{equation}
\label{eq:velocity}
    \bm{v} = \frac{\Delta \bm{u}}{\Delta t}
\end{equation}

The use of the material gradient in Equation (\ref{eq:spin}) distinquishes the material spin from the commonly known spin $\bm{w}$, which is defined using the \emph{spatial} gradient instead \cite{deSouzaNeto2008computational}, and is linked to the rotational tensor $\bm{R}$ through the relation \cite{Gurtin2010mechandtdcont}

\begin{equation}
\label{eq:gurtinrotations}
    \dot{\bm{R}} = \bm{w}\cdot\bm{R} - \bm{R}\cdot{\rm Skw}\left(\dot{\bm{U}}\cdot\bm{U}^{-1}\right)
\end{equation}

\noindent where ${\rm Skw}\left(\bm{A}\right)$ denotes the skew symmetric part of a tensor $\bm{A}$. A comparison between Equations (\ref{eq:gurtinrotations}) and (\ref{eq:Qtimederivative}) reveals the simplifications made in the use of $\bm{Q}$ instead of $\bm{R}$, that is namely the exchange of spatial spin for material spin and neglect of the correction term removing the rotated skew symmetric parts of stretch from the result. While these simplifications introduce errors in physical description of curvature, we do not find them to hinder the purpose of the regularization term, i.e., stabilization of behavior of the fictional third medium.

A linearization of Equation (\ref{eq:Qtimederivative}) with the use of exponential mapping \cite{deSouzaNeto2008computational} leads to
\begin{equation}
\label{eq:incremental_rotations}
    \bm{Q}(t_{n+1}) = {\rm exp}(\Delta t\bm{w}_X)\cdot\bm{Q}(t_n)
\end{equation}
\noindent where $t_n$ and $t_{n+1}$ are arbitrarily chosen pseudo-times of the finite element solver and $\Delta t = t_{n+1} - t_n$ is their difference. 

A logarithm can now be applied to equation (\ref{eq:incremental_rotations}) and since in the two-dimensional case, the tensors $\exp(\bm{w}_X)$ and $\bm{Q}$ commute (for 3D see \ref{app:exponentialortiz}), the rule on sum of logarithms can be applied, yielding
\begin{equation}
\label{eq:incremental_rotations_logaritmized}
    \ln \bm{Q}(t_{n+1}) = \Delta t \bm{w}_X + \ln \bm{Q}(t_n)
\end{equation}
Finally, applying the gradient operator to both sides of the equation, a simple additive incremental formulation for $\nabla \ln \bm{Q}$ is obtained:
\begin{equation}
\label{eq:gradlnRformula}
     \nabla \ln \bm{Q}(t_{n+1}) = \Delta t \nabla \bm{w}_X + \nabla \ln \bm{Q}(t_n)
\end{equation}

The material spin tensor gradient $\nabla \bm{w}_X$ can be calculated at a Gauss point of a standard second-order finite element from nodal displacements and shape functions derivatives, as detailed below. The incremental nature of the formula requires a basic implementation of material status memory, which is otherwise uncommon for hyper-elasticity, but common, e.g., for materials with plastic behavior. As an initial condition, the value $\nabla \ln \bm{Q}(t_0 = 0) = \bm{0}$ is assumed for the undeformed state of the third medium.

The definition of spatial velocity (recall Equation~(\ref{eq:velocity})) can be used, introducing the incremental displacement gradient $\bm{f} = \nabla (\Delta \bm{u})$, to move the time step length to the left side of Equation (\ref{eq:spin}):
\begin{equation}
    \Delta t\bm{w}_X = \frac{1}{2} \left(\bm{f} - \bm{f}^T\right) 
\end{equation}
To obtain the time step increment to the gradient $\bm{Q}$-rotation logarithm term (recall Equation~(\ref{eq:gradlnRformula})), only an application of the gradient operator remains:
\begin{equation}
    \Delta t\nabla\bm{w}_X = \frac{1}{2} \left(\nabla\bm{f} - \left(\nabla\bm{f}^T\right)\right) 
\end{equation}

Upon discretization of the domain into finite elements, it would be advantageous to express this relation in dependence on the nodal displacements vector~$\mathsf{d}$, its time increment $\Delta \mathsf{d}$ and their variation $\delta \mathsf{d}$ (note that $\delta \Delta\mathsf{d} = \delta \mathsf{d}$ because $\mathsf{d}(t_{n+1}) = \mathsf{d}(t_n) + \Delta \mathsf{d}$). So it follows for the incremental term and its variation that
\begin{eqnarray}
    \label{eq:spininfem}
    \Delta t \nabla \bm{w}_X &=& \mathsf{H}_w \cdot \Delta \mathsf{d}
    \\
    \delta \left(\Delta t \nabla \bm{w}_X\right) &=& \mathsf{H}_w \cdot \delta\mathsf{d}
\end{eqnarray}
\noindent where $\mathsf{H}_w$ is an appropriate linear mapping of second derivatives of element shape functions to nodal displacements. In an engineering matrix notation, Equation~(\ref{eq:spininfem}) takes on the form
\begin{equation}
\footnotesize
   \Delta t \begin{bmatrix}
    w^X_{11,1}
    \\
    w^X_{11,2}
    \\ 
    w^X_{12,1}
    \\
    w^X_{12,2}
    \\
    w^X_{21,1}
    \\
    w^X_{21,2}
    \\
    w^X_{22,1}
    \\
    w^X_{22,2}
   \end{bmatrix} = \frac{1}{2}\begin{bmatrix}
    0 & 0 & 0 & 0 & & ... & 0 & 0
    \\
    0 & 0 & 0 & 0 & & ... & 0 & 0
    \\
    N_{1,21} & -N_{1,11} & N_{2,21} & -N_{2,11} & &  ... & N_{i,21}  & -N_{i,11}
    \\
    N_{1,22} & -N_{1,12} & N_{2,22} & -N_{2,12} & &  ... & N_{i,22}  & -N_{i,12}
    \\
    -N_{1,21} & N_{1,11} & -N_{2,21} & N_{2,11} & &  ... & -N_{i,21}  & N_{i,11}
    \\
    -N_{1,22} & N_{1,12} & -N_{2,22} & N_{2,12} & &  ... & -N_{i,22}  & N_{i,12}
    \\
    0 & 0 & 0 & 0 & & ... & 0 & 0
    \\
    0 & 0 & 0 & 0 & & ... & 0 & 0
    \end{bmatrix} \begin{bmatrix}
        \Delta d_{1}
        \\
        \Delta d_{2}
        \\
        \Delta d_{3}
        \\
        \Delta d_{4}
        \\
        ...
        \\
        \Delta d_{2i-1}
        \\
        \Delta d_{2i}
    \end{bmatrix}
    \normalsize
\end{equation}

\noindent where $X_1$ and $X_2$ are components of the material coordinate tensor $\bm{X}$, the subscript $\bullet_{,k}$ denotes a derivative with regards to $X_k$, $N_1$ to $N_i$ are the shape functions of an $i$-node element dependent on the material position $\bm{X}$, and $\Delta d_1$ to $\Delta d_{2i}$ are the incremental values of displacements in the degrees of freedom corresponding to those nodes.

The relevant part of the strain energy density function, as discussed in Section~\ref{sec:model_regularization}, has a quadratic form:
\begin{equation}
    W_\mathrm{r,\ln(Q)} = \frac{1}{2}c \; \nabla \ln \bm{Q} \mathbin{\vdots} \nabla \ln \bm{Q}
\end{equation}
\noindent with $c$ being a material parameter. Its first and second variations read
\begin{eqnarray}
    \delta W_\mathrm{r,Q} &=& c\left(\nabla \ln \bm{Q} \right) \mathbin{\vdots} \delta \left(\nabla \ln \bm{Q}\right)
    \\
    \delta^2 W_\mathrm{r,Q} &=& c \; \delta \left(\nabla \ln \bm{Q} \right)\mathbin{\vdots} \delta \left(\nabla \ln \bm{Q}\right)
\end{eqnarray}
Due to the additive incremental nature of Equation (\ref{eq:gradlnRformula}) the variation $ \delta \left(\nabla \ln \bm{Q}\right)$ only depends on the variation of the material spin gradient $\nabla \bm{w}_X$ and does not depend on any material status history, i.e.,
\begin{eqnarray}
    \nabla \ln \bm{Q} (t_{n+1})  &=& \nabla \ln \bm{Q}(t_n) + \mathsf{H}_w \cdot \Delta \mathsf{d}
    \\
     \delta \left(\nabla \ln \bm{Q} \right) &=& \mathsf{H}_w \cdot \delta \mathsf{d}
\end{eqnarray}

\subsection{Alternative energy formulation with direct exponential mapping}
\label{app:exponentialortiz}

The introduction of logarithm to the $\bm{Q}$-rotation term is not strictly necessary; an alternative approach is possible using directly the exponential mapping defined in Equation (\ref{eq:incremental_rotations}). This would be especially important in a 3D implementation, where the rule on sum of logarithms (consider Equation \ref{eq:incremental_rotations_logaritmized}) cannot be used.

Upon application of the gradient operator, the expression
\begin{equation}
    \nabla \bm{Q}(t_{n+1}) = \nabla\left(\exp(\Delta t\bm{w}_X)\right)\cdot\bm{Q}(t_{n}) + \exp(\Delta t\bm{w}_X)\cdot\nabla\bm{Q}(t_{n})
\end{equation}
\noindent is obtained, with $t_n$ denoting the $n$-th time step.

Values of $\bm{Q}(t_{n})$ and $\nabla\bm{Q}(t_{n})$ are always known from the previous time step, seeded as $\bm{I}$ and $\bm{0}$, respectively, at $t_0 = 0$. For the calculation of the exponential term and its gradient, the method proposed by Ortiz~et~al.~\cite{Ortiz2001exponentialderivatives} can be applied. The exponential is expressed as a sum of an infinite series
\begin{eqnarray}
    \exp(\Delta t\bm{w}_X) &=& \sum_{k=0}^\infty \frac{1}{k!} (\Delta t\bm{w}_X)^{k}
\end{eqnarray}
\noindent The $(k+1)$-th member of this series and its gradient can be calculated from the $k$-th member and its gradient:
\begin{eqnarray}
    \exp(\Delta t\bm{w}_X)^{(k+1)} &=& \frac{1}{k+1} \exp(\Delta t\bm{w}_X)^{(k)} \cdot \left(\Delta t\bm{w}_X\right)
    \\
    \nabla \exp(\Delta t\bm{w}_X)^{(k+1)} &=& \frac{1}{k+1} \Big( \nabla \exp(\Delta t\bm{w}_X)^{(k)} \cdot \left(\Delta t\bm{w}_X\right)
    \\ \nonumber
    &+& \exp(\Delta t\bm{w}_X)^{(k)} \cdot \left(\Delta t\nabla \bm{w}_X\right) \Big)
\end{eqnarray}
\noindent considering the zeroth member as
\begin{equation}
    \exp(\Delta t\bm{w}_X)^{(0)} = \bm{I} \quad \quad \quad \nabla \exp(\Delta t\bm{w}_X)^{(0)} = \bm{0}
\end{equation}
This allows construction of the strain energy density function without the logarithm in the form
\begin{equation}
    W_\mathrm{r,Q} = \frac{1}{2}c\;\nabla \bm{Q} \mathbin{\vdots} \nabla \bm{Q}
\end{equation}

The recursive formulae used in this approach present an increase both in computational time and in implementation difficulty. Differentiation of the recursive formulas for the calculation of energy variations, i.e., forces and stiffness tangent, is certainly possible, but cumbersome. Alternatively, a finite difference scheme may be applied on the strain energy function at a further cost to computational time, as was done in our testing. In our experience, the effect of this more precise approach to penalization of curvature is negligible and does not compensate for this increase in complexity of the material model. It is thus only recommended to use this scheme in 3D context, where it is necessary.

\subsection{Penalization of Jacobian gradients}

Due to instabilities and/or stiffening arising in certain geometries of third medium elements penalized by curvature only, as discussed in Section \ref{sec:model_regularization}, enforcement of uniform volume change across an element is required. For this reason, another term is introduced into the energy density function, penalizing gradients of deformation gradient Jacobian $J$:
\begin{equation}
    W_\mathrm{r,J} = \frac{1}{2}c \; \left(\nabla J \cdot \nabla J\right)
\end{equation}
The Jacobian is defined as the determinant of the deformation gradient:
\begin{eqnarray}
    J = \det{\bm{F}}
\end{eqnarray}
\noindent and thus its gradient can be expressed as
\begin{equation}
    \nabla J = \frac{\partial \det{\bm{F}}}{\partial \bm{F}} : \nabla \bm{F}= \cof \bm{F} : \nabla \bm{F}
\end{equation}
\noindent with $\cof \bm{F} = \det(\bm{F}) \bm{F}^{-T}$ signifying the tensor cofactor operation. An efficient use can be made of an alternative definition of the tensor cofactor \cite{Bonet2016tensorcrossproduct}, leading to
\begin{eqnarray}
    J &=& \frac{1}{6}\bm{F} \times \bm{F} : \bm{F}
    \\
    \nabla J &=& \frac{1}{2} \bm{F} \times \bm{F} : \nabla \bm{F}
\end{eqnarray}
\noindent where $\times$ denotes the tensor cross product operator. In the indicial notation the expressions amount to
\begin{eqnarray}
    J &=& \frac{1}{6} \epsilon_{ikp}\epsilon_{jlq}F_{kl}F_{pq}F_{ij}
    \\
    J_{,m} &=& \frac{1}{2} \epsilon_{ikp}\epsilon_{jlq}F_{kl}F_{pq}F_{ij,m}
\end{eqnarray}
\noindent where $\epsilon_{ijk}$ is the Levi-Civita symbol and the subscript $\bullet_{,m}$ denotes the $m$-th coordinate of the gradient, i.e., a derivative with regards to the reference coordinate $X_m$.

Computing the necessary variations of the strain energy density function $W_\mathrm{r,J}$, it is now possible to derive the contributions to the first Piola-Kirchhoff stress $\bm{P}_\mathrm{r,J}$, second order stress $\bm{T}_\mathrm{r,J}$, elastic stiffness tangent $\mathbf{D}_\mathrm{r,J}$, mixed order stiffnesses $\mathbf{C}_\mathrm{PS,r,J}$ and  $\mathbf{C}_\mathrm{SP,r,J}$, and second order stiffness $\mathbf{C}_\mathrm{SS,r,J}$. In indicial notation, these can be written as
\begin{eqnarray}
    W_\mathrm{r,J} &=& \frac{1}{2}c J_{,m} J_{,m}
    \\
    P^\mathrm{r,J}_{rs} &=& c J_{,m} \frac{\partial J_{,m}}{\partial F_{rs}}
    \\
    T^\mathrm{r,J}_{rst} &=& c J_{,m} \frac{\partial J_{,m}}{\partial F_{rs,t}}
    \\
    D^\mathrm{r,J}_{rsuv} &=& c \frac{\partial J_{,m}}{\partial F_{uv}} \frac{\partial J_{,m}}{\partial F_{rs}} + c J_{,m} \frac{\partial^2 J_{,m}}{\partial F_{rs}\partial F_{uv}}
    \\
    C^\mathrm{PS,r,J}_{rsuvw} &=& c \frac{\partial J_{,m}}{\partial F_{uv,w}} \frac{\partial J_{,m}}{\partial F_{rs}} + c J_{,m} \frac{\partial^2 J_{,m}}{\partial F_{rs}\partial F_{uv,w}}
    \\
    C^\mathrm{SP,r,J}_{rstuv} &=& c \frac{\partial J_{,m}}{\partial F_{uv}} \frac{\partial J_{,m}}{\partial F_{rs,t}} + c J_{,m} \frac{\partial^2 J_{,m}}{\partial F_{rs,t}\partial F_{uv}}
    \\
    C^\mathrm{SS,r,J}_{rstuvw} &=& c \frac{\partial J_{,m}}{\partial F_{uv,w}} \frac{\partial J_{,m}}{\partial F_{rs,t}} + c J_{,m} \frac{\partial^2 J_{,m}}{\partial F_{rs,t} \partial F_{uv,w}}
\end{eqnarray}
\noindent where
\begin{eqnarray}
    \label{eq:jacobianefs1}
    \frac{\partial J_{,m}}{\partial F_{rs}} &=& \frac{1}{2} \epsilon_{irp}\epsilon_{jsq}F_{pq}F_{ij,m} + \frac{1}{2} \epsilon_{ikr}\epsilon_{jls}F_{kl}F_{ij,m} = \epsilon_{irp}\epsilon_{jsq}F_{pq}F_{ij,m}
    \\
    \frac{\partial J_{,m}}{\partial F_{rs,t}} &=& \frac{1}{2} \epsilon_{rkp}\epsilon_{slq}F_{kl}F_{pq}\delta_{mt}
    \\
    \frac{\partial^2 J_{,m}}{\partial F_{rs}\partial F_{uv}} &=& \frac{1}{2} \epsilon_{iru}\epsilon_{jsv}F_{ij,m} + \frac{1}{2} \epsilon_{iur}\epsilon_{jvs}F_{ij,m} = \epsilon_{iru}\epsilon_{jsv}F_{ij,m}
    \\
    \frac{\partial^2 J_{,m}}{\partial F_{rs}\partial F_{uv,w}} &=&  \frac{1}{2} \epsilon_{urp}\epsilon_{vsq}F_{pq}\delta_{mw} + \frac{1}{2} \epsilon_{ukr}\epsilon_{vls}F_{kl}\delta_{mw} = \epsilon_{urp}\epsilon_{vsq}F_{pq}\delta_{mw}
    \\
    \frac{\partial^2 J_{,m}}{\partial F_{rs,t}\partial F_{uv}} &=&  \frac{1}{2} \epsilon_{rup}\epsilon_{svq}F_{pq}\delta_{mt} + \frac{1}{2} \epsilon_{rku}\epsilon_{slv}F_{kl}\delta_{mt} = \epsilon_{rup}\epsilon_{svq}F_{pq}\delta_{mt}
    \\
    \label{eq:jacobianefs7}
    \frac{\partial^2 J_{,m}}{\partial F_{rs,t} \partial F_{uv,w}} &=& 0
\end{eqnarray}
\noindent with $\delta_{ij}$ as the Kronecker delta symbol.

With these first and second order stresses and stiffnesses, the first and second variation of $W_\mathrm{reg,J}$ can be constructed as
\begin{eqnarray}
    \delta W_\mathrm{r,J} &=& \bm{P}_\mathrm{r,J}:\delta \bm{F} + \bm{T}_\mathrm{r,J}:\delta \nabla \bm{F}
    \\
    \delta^2 W_\mathrm{r,J} &=&  \delta \bm{F} : \mathbf{D}_\mathrm{r,J} : \delta \bm{F} + \delta \nabla \bm{F} : \mathbf{C}_\mathrm{PT,r,J} : \delta \bm{F}
    \\ \nonumber
    &+& \delta \bm{F} : \mathbf{C}_\mathrm{TP,r,J} : \delta \nabla \bm{F} + \delta \nabla\bm{F}:\mathbf{C}_\mathrm{TT,r,J} : \delta \nabla\bm{F}
\end{eqnarray}
\noindent leading after finite element discretization and transfer to engineering notation to

\begin{eqnarray}
    \delta W_\mathrm{r,J} &=& \mathsf{P}_\mathrm{r,J}\mathsf{B} \delta \mathsf{d} + \mathsf{T}_\mathrm{r,J}\; \mathsf{H} \delta \mathsf{d}
    \\
    \delta^2 W_\mathrm{r,J} &=&  \delta \mathsf{d}^\mathrm{T} \mathsf{B}^\mathrm{T} \;\mathsf{D}_\mathrm{r,J}\; \mathsf{B} \delta \mathsf{d} + \delta \mathsf{d}^\mathrm{T}\mathsf{H}^\mathrm{T} \;\mathsf{C}_\mathrm{PT,r,J}\; \mathsf{B} \delta \mathsf{d}
    \\ \nonumber
    &+& \delta \mathsf{d}^\mathrm{T}\mathsf{B}^\mathrm{T} \;\mathsf{C}_\mathrm{TP,r,J}\; \mathsf{H} \delta \mathsf{d} + \delta \mathsf{d}^\mathrm{T}\mathsf{H}^\mathrm{T} \;\mathsf{C}_\mathrm{TT,r,J}\; \mathsf{H} \delta \mathsf{d}
\end{eqnarray}
\noindent with $\mathsf{B}$ a linear mapping of element shape function derivatives to nodal displacements and $\mathsf{H}$ likewise a linear mapping of second derivatives of element shape functions to nodal displacements.

This approach provides a universal derivation valid in all cases. It is, however, notable that due to the tensor cross product being only defined in three dimensions, a two-dimensional deformation gradient tensors have to be brought to the three-dimensional space (and later back) to allow for use of the formulas. Particularly in a plane strain case, a large simplification is possible, if a direct definition for a determinant of a second order tensor is used:
\begin{eqnarray}
    J = F_{11}F_{22} - F_{12}F_{21}
\end{eqnarray}
 Now it is possible to simply differentiate for the gradient:
\begin{eqnarray}
    J_{,m} = F_{11,m}F_{22} + F_{11}F_{22,m} - F_{12,m}F_{21} - F_{12}F_{21,m}
\end{eqnarray}
Equations (\ref{eq:jacobianefs1})-(\ref{eq:jacobianefs7}) thus greatly simplify to:
\begin{eqnarray}
    \frac{\partial J_{,m}}{\partial F_{rs}} &=& F_{11,m}\delta_{2r}\delta_{2s} + \delta_{1r}\delta_{1s}F_{22,m} - F_{12,m}\delta_{2r}\delta_{1s} - \delta_{1r}\delta_{2s}F_{21,m}
    \\
    \frac{\partial J_{,m}}{\partial F_{rs,t}} &=& \delta_{1r}\delta_{1s}\delta_{mt}F_{22} + F_{11}\delta_{2r}\delta_{2s}\delta_{mt}
    \\ \nonumber
    &-& \delta_{1r}\delta_{2s}\delta_{mt}F_{21} - F_{12}\delta_{2r}\delta_{1s}\delta_{mt}
    \\
    \frac{\partial^2 J_{,m}}{\partial F_{rs}\partial F_{uv}} &=& 0
    \\
    \frac{\partial^2 J_{,m}}{\partial F_{rs}\partial F_{uv,w}} &=& \delta_{1u}\delta_{1v}\delta_{mw}\delta_{2r}\delta_{2s} + \delta_{1r}\delta_{1s}\delta_{2u}\delta_{2v}\delta_{mw} 
    \\ \nonumber
    &-& \delta_{1u}\delta_{2v}\delta_{mw}\delta_{2r}\delta_{1s} - \delta_{1r}\delta_{2s}\delta_{2u}\delta_{1v}\delta_{mw}
    \\
    \frac{\partial^2 J_{,m}}{\partial F_{rs,t}\partial F_{uv}} &=& \delta_{1r}\delta_{1s}\delta_{mt}\delta_{2u}\delta_{2v} + \delta_{1u}\delta_{1v}\delta_{2r}\delta_{2s}\delta_{mt}
    \\ \nonumber
    &-& \delta_{1r}\delta_{2s}\delta_{mt}\delta_{2u}\delta_{1v} - \delta_{1u}\delta_{2v}\delta_{2r}\delta_{1s}\delta_{mt}
    \\
    \frac{\partial^2 J_{,m}}{\partial F_{rs,t} \partial F_{uv,w}} &=& 0
\end{eqnarray}
Note that in this plane strain case, all the tangent contributions are generally independent of the in-plane members of the deformation gradient, making them constant.

\clearpage
\section{Identification of material parameters for the silicone rubber used in physical experiment}
\label{app:matparams}

In Section \ref{sec:metamaterial}, we present a verification of the proposed third medium material model against a physical experiment of pneumatic actuation of a pattern-transforming metamaterial. The experimental sample was cast from silicone rubber \cite{havel2019p20silicone} in a 3D-printed mold. To match simulation and experimental data, material properties need to be found for this silicone rubber material.

The chosen material model is the neo-Hookean hyperelastic model with the following strain energy density

\begin{equation}
     W_\mathrm{NH,bulk} = \frac{K}{2}\ln^2{J}
    + \frac{G}{2}\left(J^{-2/3}I_1 - 3\right)
\label{eq:bulkenergyappendix}
\end{equation}

\noindent with $K$ and $G$ being the material parameters that need to be determined. They are a representation of (and when linearized around the reference configuration they exactly correspond to) the bulk and shear moduli, respectively, of linear small-strain elasticity. Consequently, they can be equivalently expressed in terms of Young's modulus $E$ and the initial Poisson's ratio $\nu$:

\begin{equation}
    K = \frac{E}{3(1-2\nu)} \quad \quad G = \frac{E}{2(1+\nu)}
\label{eq:moduli}
\end{equation}

Polymer-based elastic materials such as silicone rubbers are generally assumed to be incompressible with a Poisson's ratio approaching $\nu = 0.5$ \cite{Mott2009Poissonrationvalues}. Introducing an assumption of near incompressibility into our model and setting $\nu=0.499$, we are left with two parameters which both are just a linear function of an unknown Young's modulus $E$. No specific material treatment is required for this value of $\nu$.

\begin{figure}
    \centering
    \begin{tabular}{c c c c}
     \includegraphics[height=.25\textheight]{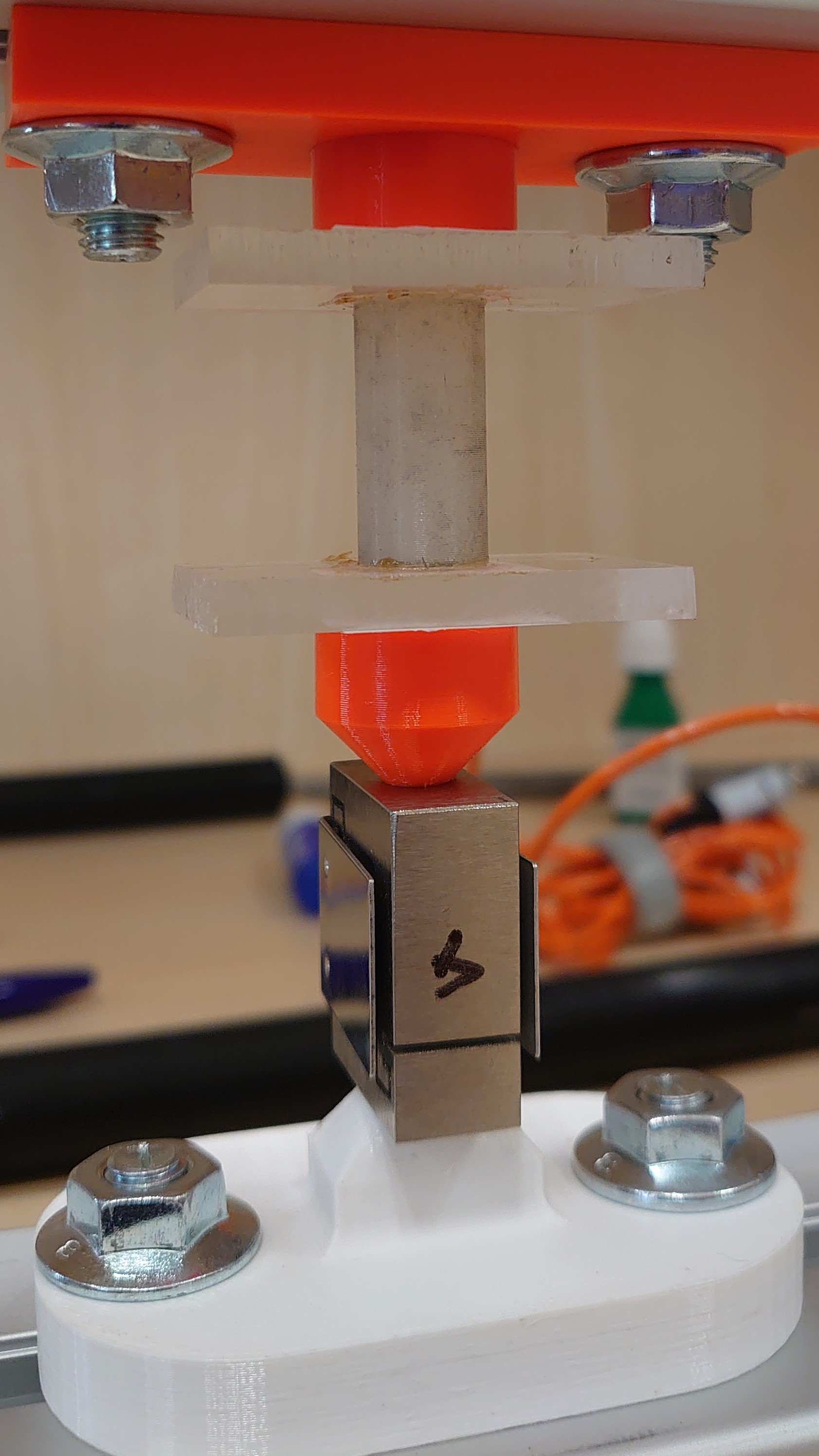}
     & 
     \includegraphics[height=.25\textheight]{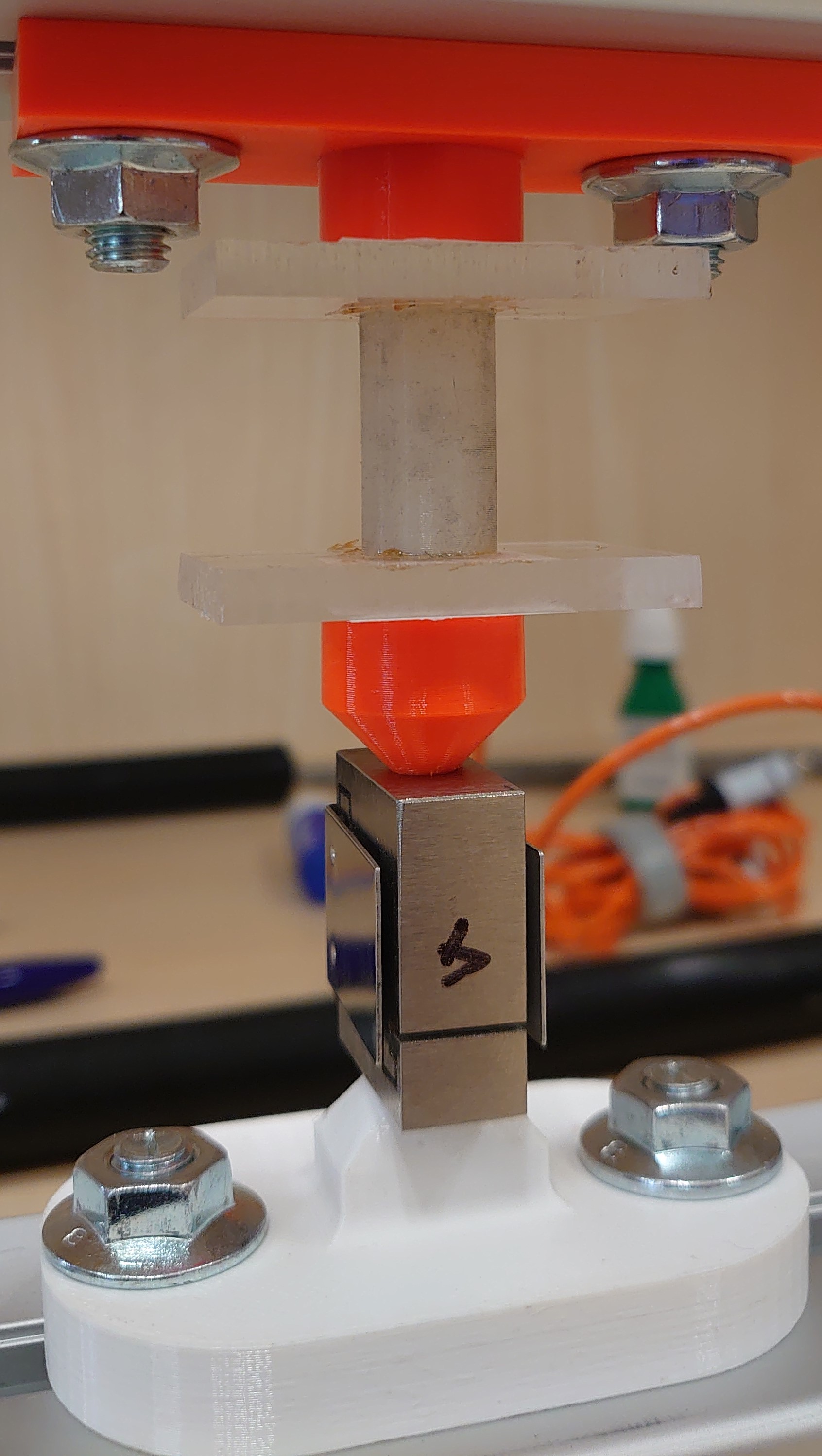}
     &
     \includegraphics[height=.25\textheight]{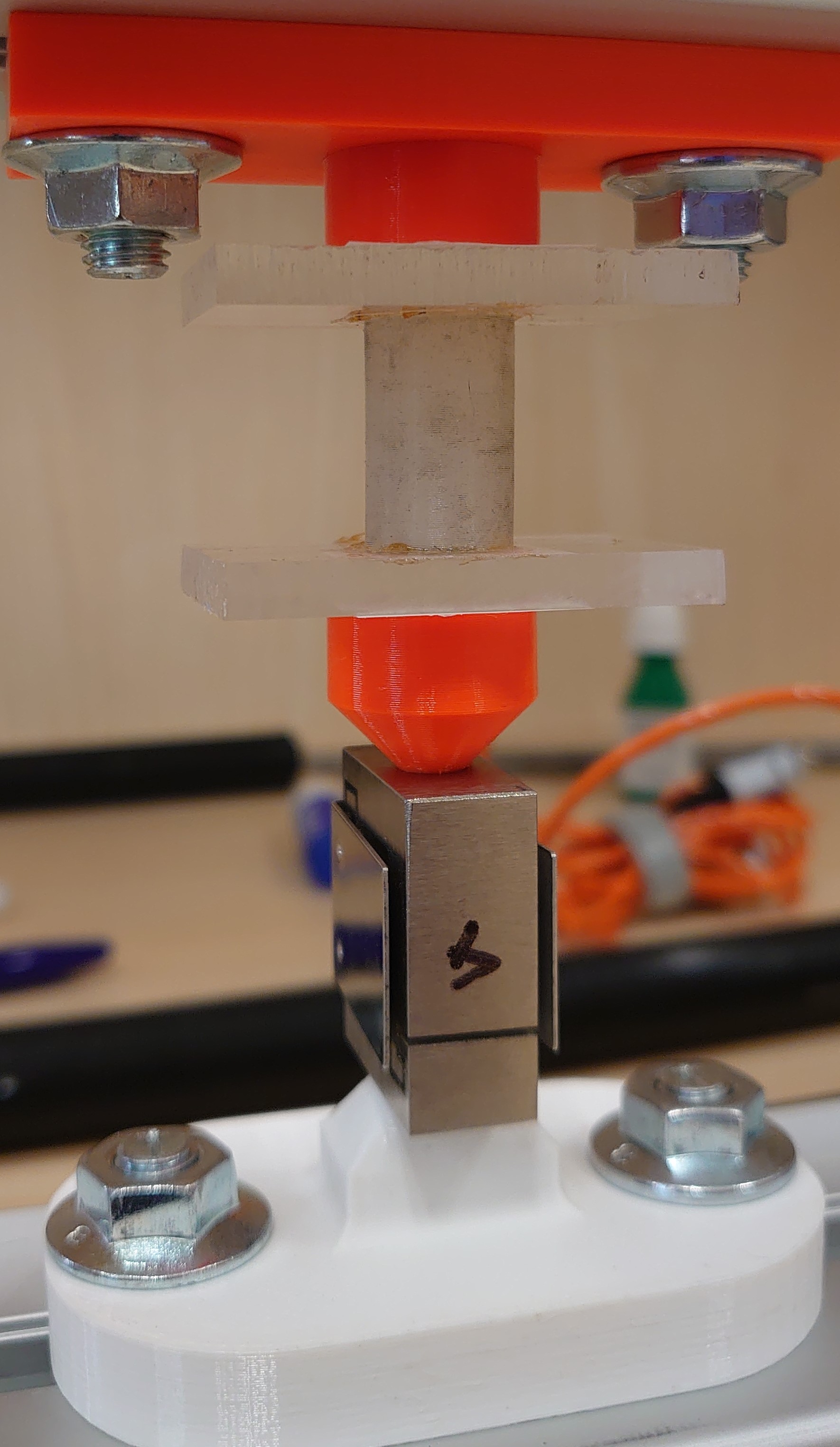}
     &
     \includegraphics[height=.25\textheight]{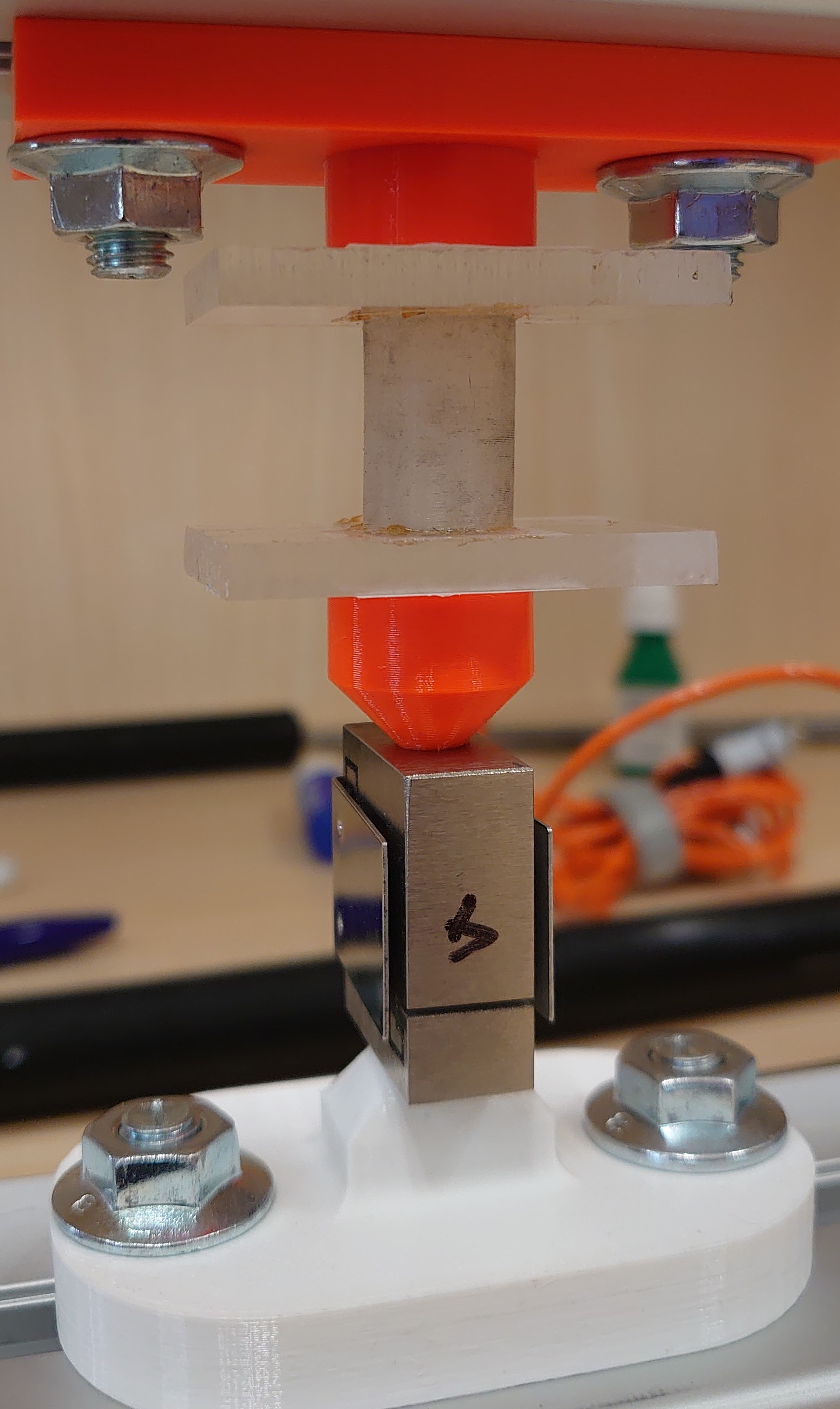}
    \end{tabular}
    \caption{Uniaxial compression of silicone rubber cylinders, various stages of loading (compressive strain increases left to right). The samples are fixed in the loading machine between lubricated plexiglass plates.}
    \label{fig:axialtests}
\end{figure}

\begin{figure}
    \centering
    \includegraphics[width=.7\textwidth]{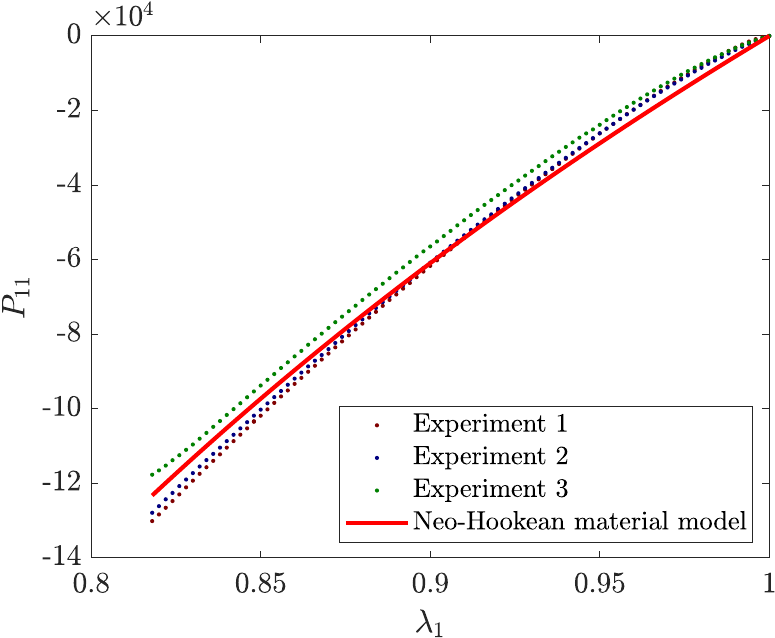}
    \caption{Stress-strain diagrams for uniaxial compression tests on cylindrical silicone rubber samples shown in Figure \ref{fig:axialtests}. Experimental data from three independent tests and a response curve of a near incompressible neo-Hookean material model with parameters determined from a least-square fit procedure.}
    \label{fig:curvefit}
\end{figure}

To determine this unknown value, uniaxial compression tests were conducted on samples of the silicone rubber material. Three separate cyllindrical samples of length $L = \SI{25}{\milli\meter}$ and diameter $d = \SI{13}{\milli\meter}$ were compressed in the axial direction in an open-hardware Thymos\footnote{\url{http://thymos.cz/}} loading frame. The experimental setup with different stages of loading can be seen in Figure \ref{fig:axialtests}. The loading continued up to the point at which each individual sample slid out of its fixings. The force-displacement curve was measured at displacement increments of $\Delta u = \SI{0.05}{\milli\meter}$. Converting the resulting data to axial stretch $\lambda_1 = (L-u)/L$ and the normal component of the first Piola-Kirchhoff stress $P_{11} = N/(\pi d^2/4)$ where $N$ is the measured normal force response, gives three stress-strain curves shown in Figure \ref{fig:curvefit} (dotted lines).

The conditions of the uniaxial compression test can be described by the neo-Hookean model. Starting from the strain energy density in Equation (\ref{eq:bulkenergyappendix}) and differentiating with respect to the deformation gradient, the first Piola-Kirchhoff stress is obtained in the indicial notation as

\begin{equation}
    P_{ij} =\frac{\partial W_\mathrm{NH,bulk}}{\partial F_{ij}} =  K \ln{J} F^{-1}_{ji} - \frac{G}{3}J^{-2/3}F^{-1}_{ji}\;I_1 + GJ^{-2/3}F_{ij}
\label{eq:NH1PK}
\end{equation}

\noindent Assuming laterally unconstrained uniaxial compression, the matrix representation of the deformation gradient takes the form of

\begin{equation}
    \mathsf{F} = \begin{bmatrix}
        \lambda_1 & 0 & 0 \\ 0 & \lambda_2 & 0 \\ 0 & 0 & \lambda_3
    \end{bmatrix}
\end{equation}

\noindent where $\lambda_i$ are the principal stretches with $\lambda_1$ oriented along the axis of the cylinder. The invariants $J$ and $I_1$ then read

\begin{equation}
    J = \lambda_1 \lambda_2 \lambda_3  \quad \quad I_1 = \lambda_1^2 + \lambda_2^2 + \lambda_3^2
\end{equation}

\noindent Substituting into Equation (\ref{eq:NH1PK}), normal components of the first Piola-Kirchhoff stress in the direction of $\lambda_i$ can be expressed as

\begin{eqnarray}
\label{eq:1PKinStretches}
    P_{ii} &=& \frac{K}{\lambda_i} \left( \ln{\lambda_1} + \ln{\lambda_2} + \ln{\lambda_3}\right) 
    \\ \nonumber
    &-& \frac{G}{3}\left(\lambda_1\lambda_2\lambda_3\right)^{-2/3}\left(\frac{1}{\lambda_i}\left(\lambda_1^2 + \lambda_2^2 + \lambda_3^2\right) - 3\lambda_i\right)
\end{eqnarray}

\noindent Thus it is possible to compute for a given axial stretch $\lambda_1$ the corresponding lateral stretches $\hat{\lambda} = \lambda_2 = \lambda_3$, using the condition of zero normal stress in the lateral direction
\begin{equation}
    P_{22} = P_{33} = 0 =\frac{K}{\hat{\lambda}} \left( \ln{\lambda_1} + 2\ln{\hat{\lambda}}\right) -\frac{G}{3}\left(\lambda_1\hat{\lambda}^2\right)^{-2/3}\left(\frac{1}{\hat{\lambda}}\left(\lambda_1^2 + 2\hat{\lambda}^2\right) - 3\hat{\lambda}\right)
\end{equation}
\noindent and then substitute back into Equation (\ref{eq:1PKinStretches}) to obtain the normal component of the first Piola-Kirchhoff stress $P_{11}$.

The curve computed as described above has been fitted to the experimental data using a least square fit procedure for the value of $E$. Errors with respect to all three experimental datasets have been weighted equally. The resulting fit is pictured in Figure \ref{fig:curvefit}. The optimal Young's modulus value found is $E = \SI{0.547}{\mega\pascal}$, which substituting into Equations (\ref{eq:moduli}) gives the material parameters of
\begin{equation}
    K = \SI{91.111}{\mega\pascal} \quad \quad G = \SI{0.182}{\mega\pascal}
\end{equation}

%% file: main.bbl
\begin{thebibliography}{10}
\expandafter\ifx\csname url\endcsname\relax
  \def\url#1{\texttt{#1}}\fi
\expandafter\ifx\csname urlprefix\endcsname\relax\def\urlprefix{URL }\fi
\expandafter\ifx\csname href\endcsname\relax
  \def\href#1#2{#2} \def\path#1{#1}\fi

\bibitem{Ren2018}
X.~Ren, R.~Das, P.~Tran, T.~D. Ngo, Y.~M. Xie, Auxetic metamaterials and structures: A review, Smart Materials and Structures 27~(2) (2018) 023001.
\newblock \href {https://doi.org/10.1088/1361-665X/aaa61c} {\path{doi:10.1088/1361-665X/aaa61c}}.

\bibitem{Ramakrishna2005emmetamatreview}
S.~A. Ramakrishna, Physics of negative refractive index materials, Reports on Progress in Physics 68 (2005) 449--521.
\newblock \href {https://doi.org/10.1088/0034-4885/68/2/R06} {\path{doi:10.1088/0034-4885/68/2/R06}}.

\bibitem{Lee2012micronanometamatreview}
J.-H. Lee, J.~P. Singer, E.~L. Thomas, Micro-/nanostructured mechanical metamaterials, Advanced {M}aterials 24~(36) (2012) 4782--4810.
\newblock \href {https://doi.org/10.1002/adma.201201644} {\path{doi:10.1002/adma.201201644}}.

\bibitem{Lakes1987negativepoisson}
R.~Lakes, Foam structures with a negative {P}oisson's ratio, Science 235 (1987) 1038--1040.
\newblock \href {https://doi.org/10.1126/science.235.4792.1038} {\path{doi:10.1126/science.235.4792.1038}}.

\bibitem{yu2018metamatreview}
X.~Yu, J.~Zhou, H.~Liang, Z.~Jiang, L.~Wu, Mechanical metamaterials associated with stiffness, rigidity and compressibility: A brief review, Progress in Materials Science 94 (2018) 114--173.
\newblock \href {https://doi.org/10.1016/j.pmatsci.2017.12.003} {\path{doi:10.1016/j.pmatsci.2017.12.003}}.

\bibitem{Goswami2019}
D.~Goswami, S.~Liu, A.~Pal, L.~G. Silva, R.~V. Martinez, {3D-Architected} soft machines with topologically encoded motion, Advanced Functional Materials 29~(24) (2019) 1808713.
\newblock \href {https://doi.org/10.1002/adfm.201808713} {\path{doi:10.1002/adfm.201808713}}.

\bibitem{Yang2015Bertoldigripper}
D.~Yang, B.~Mosadegh, A.~Ainla, B.~Lee, F.~Khashai, Z.~Suo, K.~Bertoldi, G.~M. Whitesides, Buckling of elastomeric beams enables actuation of soft machines, Advanced Materials 27 (2015) 6323--6327.
\newblock \href {https://doi.org/10.1002/adma.201503188} {\path{doi:10.1002/adma.201503188}}.

\bibitem{papka1999honeycombcrushing}
S.~D. Papka, S.~Kyriakides, Biaxial crushing of honeycombs: {P}art {I}: {E}xperiments, International Journal of Solids and Structures 36~(29) (1999) 4367--4396.
\newblock \href {https://doi.org/10.1016/S0020-7683(98)00224-8} {\path{doi:10.1016/S0020-7683(98)00224-8}}.

\bibitem{papka1999honeycombcrushinganalysis}
S.~D. Papka, S.~Kyriakides, In-plane biaxial crushing of honeycombs: {P}art {I}{I}: {A}nalysis, International Journal of Solids and Structures 36~(29) (1999) 4397--4423.
\newblock \href {https://doi.org/10.1016/S0020-7683(98)00225-X} {\path{doi:10.1016/S0020-7683(98)00225-X}}.

\bibitem{Ohno2002homogenizationonhoneycombs}
N.~Ohno, D.~Okumura, H.~Noguchi, Microscopic symmetric bifurcation condition of cellular solids based on a homogenization theory of finite deformation, Journal of the Mechanics and Physics of Solids 50 (2002) 1125--1153.
\newblock \href {https://doi.org/10.1016/S0022-5096(01)00106-5} {\path{doi:10.1016/S0022-5096(01)00106-5}}.

\bibitem{Mullin2007patterning}
T.~Mullin, S.~Deschanel, K.~Bertoldi, M.~C. Boyce, Pattern transformation triggered by deformation, Physical Review Letters 99~(8) (2007) 084301.
\newblock \href {https://doi.org/10.1103/PhysRevLett.99.084301} {\path{doi:10.1103/PhysRevLett.99.084301}}.

\bibitem{Bertoldi2010auxeticityinpatterning}
K.~Bertoldi, P.~M. Reis, S.~Willshaw, T.~Mullin, Negative {P}oisson's ratio behavior induced by an elastic instability, Advanced Materials 22 (2010) 361--366.
\newblock \href {https://doi.org/10.1002/adma.200901956} {\path{doi:10.1002/adma.200901956}}.

\bibitem{Florijn2014}
B.~Florijn, C.~Coulais, M.~V. Hecke, Programmable mechanical metamaterials, Physical Review Letters 113~(17) (2014) 175503.
\newblock \href {https://doi.org/10.1103/PhysRevLett.113.175503} {\path{doi:10.1103/PhysRevLett.113.175503}}.

\bibitem{Xiao2020activereview}
S.~Xiao, T.~Wang, T.~Liu, C.~Zhou, X.~Jiang, J.~Zhang, Active metamaterials and metadevices: A review, Journal of Physics D: Applied Physics 53~(50) (2020) 503002.
\newblock \href {https://doi.org/10.1088/1361-6463/abaced} {\path{doi:10.1088/1361-6463/abaced}}.

\bibitem{su2020dualbandpneumaticabsorber}
X.~Su, C.~Feng, Y.~Zeng, H.~Yu, A dual-band tunable metamaterial absorber based on pneumatic actuation mechanism, Optics Communications 459 (2020) 124885.
\newblock \href {https://doi.org/10.1016/j.optcom.2019.124885} {\path{doi:10.1016/j.optcom.2019.124885}}.

\bibitem{khodasevych2012reconfigurablefishnet}
I.~Khodasevych, W.~Rowe, A.~Mitchell, Reconfigurable fishnet metamaterial using pneumatic actuation, Progress in Electromagnetics Research B 38 (2012) 57--70.
\newblock \href {https://doi.org/doi.org/10.2528/PIERB11102505} {\path{doi:doi.org/10.2528/PIERB11102505}}.

\bibitem{Matia2023pressureinsoftroboticactuators}
Y.~Matia, G.~H. Kaiser, R.~F. Shepherd, A.~D. Gat, N.~Lazarus, K.~H. Petersen, Harnessing nonuniform pressure distributions in soft robotic actuators, Advanced Intelligent Systems 2023 (2023) 2200330.
\newblock \href {https://doi.org/10.1002/aisy.202200330} {\path{doi:10.1002/aisy.202200330}}.

\bibitem{Chen2018pneumaticpatterns}
Y.~Chen, L.~Jin, Geometric role in designing pneumatically actuated pattern-transforming metamaterials, Extreme Mechanics Letters 23 (2018) 55--66.
\newblock \href {https://doi.org/10.1016/j.eml.2018.08.001} {\path{doi:10.1016/j.eml.2018.08.001}}.

\bibitem{bendsoe2004TopologyOptimization}
M.~P. Bendsøe, O.~Sigmund, Topology Optimization, Springer Berlin Heidelberg, 2004.
\newblock \href {https://doi.org/10.1007/978-3-662-05086-6} {\path{doi:10.1007/978-3-662-05086-6}}.

\bibitem{Coulais2018selfcontactpathways}
C.~Coulais, A.~Sabbadini, F.~Vink, M.~van Hecke, Multi-step self-guided pathways for shape-changing metamaterials, Nature 561 (2018) 512--515.
\newblock \href {https://doi.org/10.1038/s41586-018-0541-0} {\path{doi:10.1038/s41586-018-0541-0}}.

\bibitem{hammer2000topoptusingforces}
V.~B. Hammer, N.~Olhoff, Topology optimization of continuum structures subjected to pressure loading, Structural and Multidisciplinary Optimization 19 (2000) 85--92.
\newblock \href {https://doi.org/10.1007/s001580050088} {\path{doi:10.1007/s001580050088}}.

\bibitem{Caasenbrood2020pneumatictopology}
B.~Caasenbrood, A.~Pogromsky, H.~Nijmeijer, A computational design framework for pressure-driven soft robots through nonlinear topology optimization, in: 2020 3rd IEEE International Conference on Soft Robotics (RoboSoft), IEEE, 2020, pp. 633--638.
\newblock \href {https://doi.org/10.1109/RoboSoft48309.2020.9116010} {\path{doi:10.1109/RoboSoft48309.2020.9116010}}.

\bibitem{Wriggers2006contactbook}
P.~Wriggers, Computational Contact Mechanics, Springer Berlin Heidelberg, 2006.
\newblock \href {https://doi.org/10.1007/978-3-540-32609-0} {\path{doi:10.1007/978-3-540-32609-0}}.

\bibitem{Dev2023Electroactivefreespace}
C.~Dev, G.~Stankiewicz, P.~Steinmann, On the influence of free space in topology optimization of electro-active polymers, Structural and Multidisciplinary Optimization 66 (8 2023).
\newblock \href {https://doi.org/10.1007/s00158-023-03634-5} {\path{doi:10.1007/s00158-023-03634-5}}.

\bibitem{Wriggers2013thirdmedium}
P.~Wriggers, J.~Schröder, A.~Schwarz, A finite element method for contact using a third medium, Computational Mechanics 52 (2013) 837--847.
\newblock \href {https://doi.org/10.1007/s00466-013-0848-5} {\path{doi:10.1007/s00466-013-0848-5}}.

\bibitem{Pagano2008fictdomain}
S.~Pagano, P.~Alart, Self-contact and fictitious domain using a difference convex approach, International Journal for Numerical Methods in Engineering 75 (2008) 29--42.
\newblock \href {https://doi.org/10.1002/nme.2241} {\path{doi:10.1002/nme.2241}}.

\bibitem{bog2015tmcontactwithbarrier}
T.~Bog, N.~Zander, S.~Kollmannsberger, E.~Rank, Normal contact with high order finite elements and a fictitious contact material, Computers \& Mathematics with Applications 70~(7) (2015) 1370--1390.
\newblock \href {https://doi.org/10.1016/j.camwa.2015.04.020} {\path{doi:10.1016/j.camwa.2015.04.020}}.

\bibitem{Kruse2018isogeomthirdmedium}
R.~Kruse, N.~Nguyen-Thanh, P.~Wriggers, L.~D. Lorenzis, Isogeometric frictionless contact analysis with the third medium method, Computational Mechanics 62 (2018) 1009--1021.
\newblock \href {https://doi.org/10.1007/s00466-018-1547-z} {\path{doi:10.1007/s00466-018-1547-z}}.

\bibitem{Lorez2024Euleriancontact}
F.~Lorez, M.~Pundir, D.~S. Kammer, Eulerian framework for contact between solids represented as phase fields, Computer Methods in Applied Mechanics and Engineering 418 (2024) 116497.
\newblock \href {https://doi.org/10.1016/j.cma.2023.116497} {\path{doi:10.1016/j.cma.2023.116497}}.

\bibitem{Bluhm2021contact}
G.~L. Bluhm, O.~Sigmund, K.~Poulios, Internal contact modeling for finite strain topology optimization, Computational Mechanics 67~(4) (2021) 1099--1114.
\newblock \href {https://doi.org/10.1007/s00466-021-01974-x} {\path{doi:10.1007/s00466-021-01974-x}}.

\bibitem{Frederiksen2023topologyopt}
A.~H. Frederiksen, O.~Sigmund, K.~Poulios, Topology optimization of self-contacting structures, Computational Mechanics (2023).
\newblock \href {https://doi.org/10.1007/s00466-023-02396-7} {\path{doi:10.1007/s00466-023-02396-7}}.

\bibitem{Bonet2016nonlinearcontinuumfem}
J.~Bonet, A.~J. Gil, R.~D. Wood, Nonlinear Solid Mechanics for Finite Element Analysis: Statics, Cambridge University Press, 2016.
\newblock \href {https://doi.org/10.1017/CBO9781316336144} {\path{doi:10.1017/CBO9781316336144}}.

\bibitem{Kruzik2019numericalmethods}
M.~Kružík, T.~Roubíček, Mathematical Methods in Continuum Mechanics of Solids, Springer International Publishing, 2019.
\newblock \href {https://doi.org/10.1007/978-3-030-02065-1} {\path{doi:10.1007/978-3-030-02065-1}}.

\bibitem{rivlin1948neohookean}
R.~S. Rivlin, Large elastic deformations of isotropic materials. {I}. {F}undamental concepts, Philosophical Transactions of the Royal Society of London. Series A, Mathematical and Physical Sciences 240~(822) (1948) 459--490.
\newblock \href {https://doi.org/10.1098/rsta.1948.0002} {\path{doi:10.1098/rsta.1948.0002}}.

\bibitem{Pence2015compressiblehyperelasticity}
T.~J. Pence, K.~Gou, On compressible versions of the incompressible neo-{H}ookean material, Mathematics and Mechanics of Solids 20 (2015) 157--182.
\newblock \href {https://doi.org/10.1177/1081286514544258} {\path{doi:10.1177/1081286514544258}}.

\bibitem{Smith2019largerotationgradients}
B.~Smith, F.~D. Goes, T.~Kim, Analytic eigensystems for isotropic distortion energies, ACM Transactions on Graphics 38 (2 2019).
\newblock \href {https://doi.org/10.1145/3241041} {\path{doi:10.1145/3241041}}.

\bibitem{Poya2023largerotationgradients}
R.~Poya, R.~Ortigosa, A.~J. Gil, Variational schemes and mixed finite elements for large strain isotropic elasticity in principal stretches: Closed-form tangent eigensystems, convexity conditions, and stabilised elasticity, International Journal for Numerical Methods in Engineering 124 (2023) 3436--3493.
\newblock \href {https://doi.org/10.1002/nme.7254} {\path{doi:10.1002/nme.7254}}.

\bibitem{Fischer2011IGA}
P.~Fischer, M.~Klassen, J.~Mergheim, P.~Steinmann, R.~Müller, Isogeometric analysis of 2d gradient elasticity, Computational Mechanics 47 (2011) 325--334.
\newblock \href {https://doi.org/10.1007/s00466-010-0543-8} {\path{doi:10.1007/s00466-010-0543-8}}.

\bibitem{Horak2020gradientpolyconvexity}
M.~Horák, M.~Kružík, Gradient polyconvex material models and their numerical treatment, International Journal of Solids and Structures 195 (2020) 57--65.
\newblock \href {https://doi.org/10.1016/j.ijsolstr.2020.03.006} {\path{doi:10.1016/j.ijsolstr.2020.03.006}}.

\bibitem{geuzaine2009gmsh}
C.~Geuzaine, J.~F. Remacle, Gmsh: A {3-D} finite element mesh generator with built-in pre- and post-processing facilities, International Journal for Numerical Methods in Engineering 79 (2009) 1309--1331.
\newblock \href {https://doi.org/10.1002/nme.2579} {\path{doi:10.1002/nme.2579}}.

\bibitem{Gill1974newtonforoptimization}
P.~E. Gill, W.~Murray, Newton-type methods for unconstrained and linearly constrained optimization, Mathematical Programming 7 (1974) 311--350.
\newblock \href {https://doi.org/10.1007/BF01585529} {\path{doi:10.1007/BF01585529}}.

\bibitem{Fletcher1977modifiedNewton}
R.~Fletcher, T.~L. Freeman, A modified {N}ewton method for minimization, Journal of Optimization Theory and Applications 23 (1977) 357--372.
\newblock \href {https://doi.org/10.1007/BF00933446} {\path{doi:10.1007/BF00933446}}.

\bibitem{Crisfield2000patchtest}
M.~A. Crisfield, Re-visiting the contact patch test, International Journal for Numerical Methods in Engineering 48 (2000) 435--449.
\newblock \href {https://doi.org/10.1002/(SICI)1097-0207(20000530)48:3<435::AID-NME891>3.0.CO;2-V} {\path{doi:10.1002/(SICI)1097-0207(20000530)48:3<435::AID-NME891>3.0.CO;2-V}}.

\bibitem{Zavarise2009improvedNTS}
G.~Zavarise, L.~de~Lorenzis, A modified node-to-segment algorithm passing the contact patch test, International Journal for Numerical Methods in Engineering 79 (2009) 379--416.
\newblock \href {https://doi.org/10.1002/nme.2559} {\path{doi:10.1002/nme.2559}}.

\bibitem{deSouzaNeto2008computational}
E.~A. de~Souza~Neto, D.~Perić, D.~R.~J. Owen, Computational Methods for Plasticity: Theory and Applications, Wiley, 2008.
\newblock \href {https://doi.org/10.1002/9780470694626} {\path{doi:10.1002/9780470694626}}.

\bibitem{Gurtin2010mechandtdcont}
M.~E. Gurtin, E.~Fried, L.~Anand, The Mechanics and Thermodynamics of Continua, Cambridge University Press, 2010.
\newblock \href {https://doi.org/10.1017/CBO9780511762956} {\path{doi:10.1017/CBO9780511762956}}.

\bibitem{Ortiz2001exponentialderivatives}
M.~Ortiz, R.~A. Radovitzky, E.~A. Repetto, The computation of the exponential and logarithmic mappings and their first and second linearizations, International Journal for Numerical Methods in Engineering 52 (2001) 1431--1441.
\newblock \href {https://doi.org/10.1002/nme.263} {\path{doi:10.1002/nme.263}}.

\bibitem{Bonet2016tensorcrossproduct}
J.~Bonet, A.~J. Gil, R.~Ortigosa, On a tensor cross product based formulation of large strain solid mechanics, International Journal of Solids and Structures 84 (2016) 49--63.
\newblock \href {https://doi.org/10.1016/j.ijsolstr.2015.12.030} {\path{doi:10.1016/j.ijsolstr.2015.12.030}}.

\bibitem{havel2019p20silicone}
Havel Composites, P20 Liquid Silicone Data Sheet, revised 10 2023 (11 2019).

\bibitem{Mott2009Poissonrationvalues}
P.~H. Mott, C.~M. Roland, Limits to {P}oisson's ratio in isotropic materials, Physical Review B - Condensed Matter and Materials Physics 80~(13) (2009) 132104.
\newblock \href {https://doi.org/10.1103/PhysRevB.80.132104} {\path{doi:10.1103/PhysRevB.80.132104}}.

\end{thebibliography}
